\documentclass[sn-mathphys]{sn-jnl}

\usepackage{tensor}
\usepackage{subcaption}
\usepackage{IEEEtrantools}

\jyear{2021}%

\theoremstyle{thmstyleone}%
\theoremstyle{thmstyletwo}%

\theoremstyle{thmstylethree}%

\begin{document}

\title[\textcolor{white}{}]{Steps Towards Generalization of Tensionless String Theory with Contact Interactions as  Wilson Loop of Non-Abelian Yang-Mills Theory}


\author{\fnm{Pongwit} \sur{Srisangyingcharoen}}\email{pongwits@nu.ac.th}

\affil{\orgdiv{The Institute for Fundamental Study}, \orgname{Naresuan University}, \orgaddress{\city{Phisanulok}, \postcode{65000}, \country{Thailand}}}


\abstract{We propose a possible modification to the tensionless string model with contact interactions. The proposed model aims to reproduce the expectation value of a non-Abelian Wilson loop in Yang-Mills theory by integrating out string degrees of freedom with a fixed worldsheet boundary. To reproduce path-ordering along the worldsheet boundary, we introduce Lie algebra-valued fields on the string worldsheet, whose dynamics are determined by the topological BF action. Without bulk contributions, we show that the model describes the non-Abelian Wilson loop, neglecting the effects of self-interactions. Finally, we test the reproduction of the Wilson loop with three-point interaction in the case of $SU(2)$.}




\maketitle

\section{Introduction}\label{sec1}

Yang-Mills theories serve as important building blocks of the standard model. They are considered one of the most successful theories of fundamental particles ever tested experimentally. There are several viewpoints suggesting possible connections between Yang-Mills and string theories. Dating back to the seventies, it was known that  string theory reproduces tree diagrams of Yang-Mills theory in the limit of low energies or infinite string tension \cite{Neveu:1971mu}. This perspective proves useful
for understanding the structure of amplitude relations for both string theory and
field theory. Many structural aspects in the amplitudes of both theories were widely studied \cite{Kawai:1985xq,Stieberger:2022lss,PhysRevD.78.085011,Stieberger:2009hq,Bjerrum-Bohr:2009ulz,Cachazo:2013gna,Stieberger:2016lng,Srisangyingcharoen:2020lhx}. Another connection has been noticed by 't Hooft \cite{tHooft:1973alw} that in the large $N$ limit, a pure $SU(N)$ Yang-Mills theory diagram matches a perturbative expansion of a string theory with the string coupling constant $1/N$. 

Quite recently, the formulation of Abelian Yang-Mills theories as the tensionless limit of a spinning string with contact interactions was formulated in \cite{Edwards:2014cga} and \cite{Edwards:2014xfa}. In addition, by using the worldline approach, one can relate the theory of interacting strings to Wilson lines in spinor quantum electrodynamics in the tensionless limit \cite{Edwards:2015hka}. The idea of this formalism was initiated by Mansfield \cite{Mansfield:2011eq} which expresses an electromagnetic field strength tensor of two moving charges as a string-like object supported on a worldsheet $\Sigma$ bounded by particle worldlines expressed in the form
\begin{equation}
    F_{\mu\nu}(X)=-q\int_\Sigma \delta^4(X-Y) d\Sigma_{\mu\nu}(Y) \label{F LOF}
\end{equation}
where $d\Sigma_{\mu\nu}$ is an infinitesimal area element on the surface defined as
\begin{equation}
    d\Sigma_{\mu\nu}(X)=\epsilon^{ab}\partial_aX_\mu\partial_bX_\nu d^2\xi.\label{area ele}
\end{equation}
$X^\mu(\xi)$ is a spacetime vector with the worldsheet coordinates $\xi^a, a=1,2$ lying in the domain $\Sigma$ with boundary $\partial\Sigma$ on which $X^\mu\big\rvert_{\partial\Sigma}=\omega^\mu$ and $\partial_a$ is the derivative with respect to world-sheet co-ordinates $\xi^a$. The expression (\ref{F LOF}) satisfies the Gauss' law, $\partial_\mu F^{\mu\nu}=J^\nu$ with a current density supported on the worldsheet boundaries
\begin{equation}
    J^\mu(X)=q\oint_{\partial\Sigma} \delta^4(X-Y) dY^\mu.
\end{equation}

Substituting (\ref{F LOF}) in the pure electromagnetic action, $S=\frac{1}{4}\int d^4x F^{\mu\nu}F_{\mu\nu}$, gives
\begin{equation}
    S_\text{EM}=\int d^4x \mathcal{L}(x)=\frac{q^2}{4}\int_\Sigma d\Sigma_{\mu\nu}(X(\xi))\delta^4(X(\xi)-X(\tilde{\xi}))d\Sigma^{\mu\nu}(X(\tilde{\xi})). \label{SEM}
\end{equation}
Due to the appearance of the delta function, the action is non-vanishing when the worldsheet coordinates coincide, $\xi=\tilde{\xi}$, or when any two points coincide, i.e. $X(\xi)=X(\tilde{\xi})$. This splits (\ref{SEM}) into two pieces as
\begin{equation}
     S_\text{EM}=\frac{q^2}{4}\delta^2(0)\text{Area}(\Sigma)+\frac{q^2}{4}\int_\Sigma d\Sigma_{\mu\nu}(X(\xi))\delta^4(X(\xi)-X(\tilde{\xi}))d\Sigma^{\mu\nu}(X(\tilde{\xi}))\bigg\rvert_{\xi\neq \tilde{\xi}}. \label{string with contact action}
\end{equation}
The first piece contains the area of the worldsheet corresponding to the Nambu-Goto action of string theory, albeit with a divergent coefficient.
The latter piece provides a more interesting interpretation. It implies a contact interaction that occurs when the worldsheet intersects with itself. This is unusual from string perspectives in which the standard interactions in string theory are caused by joining and splitting worldsheets. Similar interactions have previously been discussed by Kalb and Ramond \cite{Kalb:1974yc}. From now on we denote the second term of (\ref{string with contact action}) by $S_I[X]$.

By replacing the Nambu-Goto action with the classically equivalent Polyakov  action 
\begin{equation}
    S_P[X,g]=-\frac{1}{4\pi\alpha'}\int_\Sigma d^2\xi \sqrt{g} g^{ab} \partial_a X^\mu(\xi)\partial_bX_\mu(\xi), \label{Polyakov}
\end{equation}
it can be shown perturbatively \cite{Curry:2017cnu} that the partition function of a tensionless four-dimensional string with the contact interaction $S_I$ whose worldsheet $\Sigma$ spans the closed loop $\partial\Sigma$ is similar to the Wilson loop for Abelian gauge theory associated with the closed curve $\partial\Sigma$ in flat Euclidean space at the first leading order.

To illustrate this, let us determine an expectation value of the contact interaction $\langle S_I \rangle_\Sigma$  where we defined $\langle \Omega \rangle_\Sigma$, the worldsheet average of a quantity $\Omega$ over all surfaces $\Sigma$ spanning $\partial\Sigma$, as 
\begin{equation}
    \langle \Omega\rangle_\Sigma=\frac{1}{Z_P}\int \mathcal{D}g\mathcal{D}X \Omega e^{-S_p[X,g]}\label{WS average}
\end{equation} 
with a normalization constant $Z_P$. According to the expression of $S_I$ in (\ref{string with contact action}), we apply the Fourier decomposition to the delta function to obtain
\begin{equation}
    S_I=\frac{q^2}{4}\int \frac{d^4k}{(2\pi)^4} d^2\xi d^2\xi' \ V_k^{\mu\nu}(\xi)V_{-k \ \mu\nu}(\xi') \label{SI vertex}
\end{equation}
where $V_k^{\mu\nu}(\xi)$ is the vertex operator defined as 
\begin{equation}
    V_k^{\mu\nu}(\xi)=\epsilon^{ab}\partial_a X^\mu(\xi)\partial_b X^\nu(\xi) e^{ik\cdot X(\xi)}. \label{Vertex V1}
\end{equation}
The fact that the theory lives in non-critical dimensions may concern some readers as it usually leads to the notorious Weyl anomaly. Luckily, this issue is cured by the presence of Dirac-delta function. It was shown in \cite{Mansfield:2011eq, Edwards:2014xfa} that the expectation value of the delta-function decouples from the scale of the worldsheet
metric. Therefore, this allows us not to integrate over $g_{ab}$ but rather choose a value for it.

The expression (\ref{SI vertex}) can be separated into two terms by introducing a projection operator $\mathbb{P}_k$ which is defined as
\begin{equation}
    \mathbb{P}_k(X)^\mu=X^\mu- k^\mu\frac{k\cdot X}{k^2}. \label{proj op}
\end{equation}
The operator projects any 4-vectors onto their transverse directions compared to the vector $k$. Therefore, the vertex operator now takes the form
\begin{align}
    V_k^{\mu\nu}(\xi)= \tilde{V}_k^{\mu\nu}(\xi)-\partial_a\bigg(2i\epsilon^{ab} k^{[\mu}\partial_b \mathbb{P}_k(X)^{\nu]}\frac{e^{ik\cdot X}}{k^2} \bigg)\label{Vertex V2}
\end{align} 
where $\tilde{V}_k^{\mu\nu}(\xi)$ is a projected vertex operator defined as
\begin{equation}
    \tilde{V}_k^{\mu\nu}=\epsilon^{ab}\partial_a \mathbb{P}_k(X)^\mu\partial_b \mathbb{P}_k(X)^\nu e^{ik\cdot X}. \label{projected ver}
\end{equation}
Note that as $k_\mu\mathbb{P}_k(X)^\mu=0$, the vertex operator must satisfy $k_\mu\widetilde{V}_k^{\mu\nu}=0$. Consequently, when inserting (\ref{Vertex V2}) into (\ref{SI vertex}), we find
\begin{align}
    S_I=&\frac{q^2}{4}\int \frac{d^4k}{(2\pi)^4} d^2\xi d^2\xi'\ \tilde{V}_k^{\mu\nu}(\xi)\tilde{V}_{-k \ \mu\nu}(\xi') \nonumber \\
    &+\frac{q^2}{2}\int\frac{d^4k}{(2\pi)^4}\oint_{\partial\Sigma}\oint_{\partial\Sigma} d\mathbb{P}_k(X)^\mu(\xi)\frac{e^{ik\cdot(X(\xi)-X(\xi'))}}{k^2}d\mathbb{P}_k(X)_\mu(\xi'). \label{SI2}
\end{align}
To obtain the above expression, Stoke's theorem was used.

It turns out that averaging over the worldsheet using the standard string action will suppress the first term. To see this, we use Wick's theorem to evaluate the expectation of products of fields. According to Wick's theorem for the bosonic string,
\begin{equation}
    X^\mu(\xi)X^\nu(\xi')=:X^\mu(\xi)X^\nu(\xi'):+\alpha'\delta^{\mu\nu}G(\xi,\xi') \label{Wick}
\end{equation}
$G(\xi,\xi')$ is the Green function for the worldsheet Laplacian. The colons indicate normal ordering and since the field $X$ can be expanded around the classical field $X_c$, so the expectation value of the normal ordered part is
\begin{equation}
    \langle:X^\mu(\xi)X^\nu(\xi'):\rangle_\Sigma= X_c^\mu(\xi) X_c^\nu(\xi').
\end{equation}
It is not hard to find that the expression for the projected vertex operator is
\begin{equation}
    \tilde{V}_k^{\mu\nu}=:\tilde{V}_k^{\mu\nu}:e^{-\alpha'\pi k^2 G(\xi,\xi)}. \label{projected ver2}
\end{equation}
The Green's function $G(\xi,\xi')$
can be evaluated from the heat kernel $\mathcal{G}$ via
\begin{equation}
    G(\xi,\xi')=\int_0^\infty \mathcal{G}(\xi,\xi';\tau)d\tau
\end{equation}
satisfying
\begin{equation}
    \frac{\partial}{\partial \tau}\mathcal{G}=-\triangle\mathcal{G}, \qquad \mathcal{G}(\xi,\xi';0)=\frac{1}{\sqrt{g}}\delta^2(\xi-\xi')
\end{equation}
where the Laplacian $\triangle$ is given by
\begin{equation}
    \frac{1}{\sqrt{g}}\frac{\partial}{\partial\xi^a}\Big(\sqrt{g }g^{ab}\frac{\partial}{\partial\xi^b}\Big).
\end{equation}
The general form of the heat kernel can be written using the Seeley-DeWitt expansion \cite{DeWitt:1965jb} which can be modified to a manifold with boundary \cite{McAvity:1990we}\cite{McAvity:1991xf}. If $\sigma_r(\xi,\xi')$ is twice the square of the length of the geodesic path connecting between $\xi$ and $\xi'$ with $r$ reflections at the boundary, then the heat kernel is obtained by writing 
\begin{equation}
    \mathcal{G}(\xi,\xi';\tau)=\frac{1}{4\pi\tau}\sum_r \exp{\bigg(-\frac{\sigma_r(\xi,\xi')}{2\tau}\bigg)} \Omega_r(\xi,\xi';\tau). \label{DeWitt}
\end{equation}
The function $\Omega_r$ can be expanded as a power series of $\tau$ which is
\begin{equation}
    \Omega_r(\xi,\xi';\tau)=\sum_n^\infty a^r_n(\xi,\xi')\tau^n  
\end{equation}
where $a^r_n(\xi,\xi')$ are called the Seeley-DeWitt coefficients.

According to \cite{McAvity:1990we}, for $\xi=\xi'$, the coefficients of the first few orders are evaluated as 
\begin{equation}
    a^0_0(\xi,\xi)=1, \quad a^0_1(\xi,\xi)=\frac{1}{6}R(\xi), \quad a^1_0(\xi,\xi)=-1, \quad a^1_1(\xi,\xi)=-\frac{1}{6}R(\xi)
\end{equation}
where $R$ is the Ricci scalar. It is not difficult to see from the expression (\ref{DeWitt}) that the heat kernel is divergent at a small $\tau$, thus the value of Green's function at co-incident points should be regulated with a short-distance cut-off, $\epsilon$, via an integral
\begin{equation}
    G_\epsilon(\xi,\xi)=\int_\epsilon^\infty\mathcal{G}(\xi,\xi;\tau).
\end{equation}
In this limit as $\tau\rightarrow 0$ and for $\xi\approx\xi'$, it is sufficient to obtain the asymptotic version of (\ref{DeWitt}) by including only zero and one reflection terms as
\begin{equation}
    \mathcal{G}(\xi,\xi';\tau)=\frac{1}{4\pi\tau}\bigg[ \exp{\Big(-\frac{\sigma_0}{2\tau}\Big)}-\exp{\Big(-\frac{\sigma_1}{2\tau}\Big)}\bigg].
\end{equation}
At the co-incident points, $\sigma_0=0$. Therefore, $G_\epsilon(\xi,\xi)$ reads
\begin{align}
    G_\epsilon(\xi,\xi)&\approx\int_\epsilon^\infty d\tau \ \frac{1}{4\pi\tau}\Big(1-\exp{\Big(-\frac{\sigma_1}{2\tau}\Big)}\Big) \nonumber \\
    &=\begin{cases}
    \frac{\sigma_1}{(8\pi\epsilon)}, & \sigma_1\ll \epsilon \\
    \frac{1}{4\pi}\ln\Big(\frac{\sigma_1}{\epsilon}\Big), & \sigma_1\gg \epsilon. \label{psi}
  \end{cases}
\end{align}

As a result, the Green's function at co-incident points $G(\xi,\xi)$ is zero on the worldsheet boundary and diverges as $\xi$ moves away from the boundary into the interior of the worldsheet. Since the theory was in the Euclidean signature, $k^2>0$, the projected vertex operator (\ref{projected ver2}) is suppressed inside the worldsheet for which $\alpha'k^2$ is finite. This suppression gets further amplified when taking the tensionless limit $\alpha' \rightarrow \infty$ into consideration. What remains in the expectation of $S_I$ is that of the second term in (\ref{SI2})
\begin{equation}
    \langle S_I \rangle_\Sigma=\frac{q^2}{2}\int\frac{d^4k}{(2\pi)^4}\oint_{\partial\Sigma}\oint_{\partial\Sigma} d\mathbb{P}_k(X)^\mu(\xi)\frac{e^{ik\cdot(X(\xi)-X(\xi'))}}{k^2}d\mathbb{P}_k(X)_\mu(\xi'). \label{SI3}
\end{equation}
To discard the bulk term in (\ref{SI2}), one needs to take good care of some possible divergences appearing when the vertex operators are placed close to each other either in the bulk or near the boundary of the worldsheet. A detailed analysis can be found in \cite{Edwards:2014xfa}.

The expression (\ref{SI3}) is exactly the integral over $\partial\Sigma$ of the Abelian gauge field propagator in the Lorenz gauge
\begin{equation}
    \frac{q^2}{2} \oint_{\partial\Sigma} \oint_{\partial\Sigma} d\xi_1^\mu d\xi_2^\nu \big\langle A_\mu(\xi_1)A_\nu(\xi_2) \big\rangle_{\text{YM}} .
\end{equation}
The subscript YM notifies that the expectation was made in the Abelian Yang-Mills theory. This is the first non-trivial term in the perturbative expansion of the Wilson loop
\begin{equation}
    \langle \exp{(-q\oint_{\partial\Sigma}A\cdot d\xi)} \rangle_{\text{YM}}.
\end{equation}

This suggests that the expectation value of the Wilson loop could be expressed as the worldsheet average of the exponential of $S_I$. However, a difficulty arises as divergences appear when exponentiating $S_I$ which potentially spoils the suppression. Fortunately, no such terms are produced in the supersymmetric generalization of the model. It appears that the expectation value of the super Wilson loop for (non-supersymmetric) Abelian gauge theory can be expressed as the worldsheet average of the spinning string with a contact interaction \cite{Edwards:2014cga},\cite{Edwards:2014xfa}.

In this paper, we would like to construct a suitable modification to the string model to reproduce the expectation value of a non-Abelian Wilson loop in the Yang-Mills theory. The difficulty which impedes the non-Abelian generalization is incorporating Lie algebras into the theory. This can be done by simply introducing Lie algebra-valued degrees of freedom into the worldsheet. However, one needs to find suitable dynamics to describe them. These new degrees of freedom have to generate the interaction
vertices of Yang-Mills theory and reduce to path-ordered generators on the boundary. If successfully formulated, the theory would provide an alternative channel to investigate a non-Abelian Yang-Mills theory. 

The outline of this paper is as follows. In section two, we develop the tensionless string model to incorporate a boundary path-ordering of Lie generators and at the end of the section a partition function for the non-Abelian generalization of the string model with contact interactions is proposed. In section three, the evaluation of the partition function is calculated on the worldsheet boundary. We review an effective BF theory in section four as well as compute a worldsheet gauge propagator. In section five, we evaluate a functional average of the square of string contact interactions which is partial to the partition function and compare the result with the expectation value in the $SU(2)$ gauge theory of the Wilson loop. Finally, we summarise our results in section six.


\section{Production of Boundary Path-ordering of the String Model}\label{sec2}
To generalize the string model to describe the non-Abelian Yang-Mills theory, at the very least we need to introduce Lie algebra-valued world-sheet degrees of freedom $\phi^R(\xi)$ to try to reproduce the Lie algebra structure of Yang-Mills propagators
\begin{equation}
    S_I^\phi=q^2\int_\Sigma d\Sigma_{\mu\nu}(\xi)\phi^R(\xi)\delta^4(X(\xi)-X(\tilde{\xi}))d\Sigma^{\mu\nu}(\tilde{\xi})\phi_R(\tilde{\xi}). \label{non-AB}
\end{equation}
Consequently, we need a Lagrangian to describe the dynamics of $\phi^R$. This has to be gauge-invariant to preserve the spacetime gauge invariance of the contact interaction (\ref{non-AB}) and Weyl invariant to satisfy the usual organizing
principle of string theories. In addition, the newly introduced degrees of freedom are expected to reduce to path-ordered generators on the boundary to represent a non-Abelian gauge theory. We propose that the action which meets the minimum criteria is the topological BF action \cite{horowitz1989,Blau:1989bq}, i.e.
\begin{equation}
    S_\text{BF}[\phi,\mathcal{A}]=2\int_\Sigma d^2\xi \ \epsilon^{ij} \text{tr}(\phi \mathcal{F}_{ij}) \label{BF}
\end{equation}
with the field strength tensor $\mathcal{F}_{ij}=\partial_i\mathcal{A}_j-\partial_j\mathcal{A}_i+[\mathcal{A}_i,\mathcal{A}_j]$. 
The action (\ref{BF}) has a close connection to the Yang-Mills action in two dimensions as they are equivalent in the zero coupling constant limit \cite{witten1991,Witten:1992xu}. Remember that both fields $\phi$ and $\mathcal{A}$ are Lie algebra-valued fields. They can be written in terms of a set of generators $\{T^R\}$ as $\phi=\phi_RT^R$ and $\mathcal{A}=\mathcal{A}_RT^R$.\footnote{tr($T^AT^B$)=$\frac{1}{2}\eta^{AB}$ and $[T^A,T^B]=if\indices{^A^B_C}T^C$} The worldsheet 1-form $\mathcal{A}$ is intrinsic to the worldsheet and it differs from the actual spacetime gauge field $A$ in the gauge theory whose dynamics we wish to reformulate.

It can be shown that when we define the partition function corresponding to the BF action as
\begin{align}
        Z=\frac{1}{\text{Vol}}\int D\phi D\mathcal{A} e^{-S_\text{BF}[\phi,\mathcal{A}]}\text{tr}\big(\mathcal{P}\big(e^{-\oint_{\partial\Sigma} \mathcal{A}\cdot d\xi}\big)\big),   \label{par-fn-BF}
\end{align}
one can obtain a boundary correlation function of scalar fields as path-ordering of Lie generators. A Wilson loop is inserted along the boundary of $\Sigma$. To remove all the gauge redundancy, we apply the axial gauge-fixing condition via the insertion
\begin{equation}
    1=\int D\Lambda \delta(\mathbf{n}\cdot \mathcal{A}^\Lambda) \text{det}\bigg(\frac{\delta \mathbf{n}\cdot \mathcal{A}^\Lambda}{\delta \Lambda} \bigg) \label{axial gauge}
\end{equation}
with a fixed vector $\mathbf{n}$. Therefore, the partition function (\ref{par-fn-BF}) takes the form
\begin{align}
     Z=\mathcal{N}\int D\phi D\mathcal{A}  \delta(\mathbf{n}\cdot \mathcal{A}) \text{det}(\mathbf{n}\cdot \mathcal{D}) e^{-S_\text{BF}[\phi,\mathcal{A}]}\text{tr}\big(\mathcal{P}\big(e^{-\oint_{\partial\Sigma} \mathcal{A}\cdot d\xi}\big)\big)
\end{align}
where the covariant derivative $\mathcal{D}_\mu$ is defined as $\partial_\mu+\mathcal{A}_\mu$. To obtain this, we used the fact that the integrand and the measures are gauge invariant which can then be renamed from $(\phi^\Lambda,\mathcal{A}^\Lambda)$ to $(\phi,\mathcal{A})$.

By introducing a source term for the scalar field, one can construct a generating functional as
\begin{align}
    Z[J]=\mathcal{N}\int D\phi D\mathcal{A}  \delta(\mathbf{n}\cdot \mathcal{A}) \text{det}(\mathbf{n}\cdot \mathcal{D}) e^{-S_\text{BF}[\phi,\mathcal{A}]+2\int d^2\xi\text{tr}(J\phi)}\text{tr}\big(\mathcal{P}\big(e^{-\oint_{\partial\Sigma} \mathcal{A}\cdot d\xi}\big)\big). \label{gen-fn}
\end{align}
Integrating out the field $\phi$ generates the constraint via $\delta$-function as
\begin{align}
    Z[J]=\widetilde{\mathcal{N}}\int D\mathcal{A}  \delta(\mathbf{n}\cdot \mathcal{A}) \text{det}(\mathbf{n}\cdot \mathcal{D}) \delta(\frac{1}{2}(\epsilon^{ij}\mathcal{F}_{ij}-J)) \text{tr}\big(\mathcal{P}\big(e^{-\oint_{\partial\Sigma} \mathcal{A}\cdot d\xi}\big)\big). \label{gen-fn2}
\end{align}

For simplicity, we will content ourselves to consider the worldsheet with the topology of a disk $D^2$ which is topologically equivalent to the upper half-plane. Accordingly, it is convenient to work in Cartesian coordinates, hence, (\ref{gen-fn2}) taking the form
\begin{equation}
            Z[J]=\widetilde{\mathcal{N}}\int D\mathcal{A}  \delta(\mathcal{A}_y) \text{det}( \mathcal{D}_y) \delta(\mathcal{F}-\frac{1}{2}J) \text{tr}\big(\mathcal{P}\big(e^{-\int \mathcal{A}_x dx}\big)\big). \label{gen-fn3}
\end{equation}
where $\mathcal{F}=\partial_x\mathcal{A}_y-\partial_y\mathcal{A}_x+[\mathcal{A}_x,\mathcal{A}_y]$ and we chose  the reference vector $\mathbf{n}$ to be a unit vector pointing in $y$-direction. 

After integrating out $\mathcal{A}_y$, the generating functional (\ref{gen-fn3}) becomes
\begin{equation}
            Z[J]=\widetilde{\mathcal{N}}\int D\mathcal{A}_x  \text{det}( \partial_y) \delta(\partial_y \mathcal{A}_x+\frac{1}{2}J) \text{tr}\big(\mathcal{P}\big(e^{-\int \mathcal{A}_x dx}\big)\big). 
\end{equation}
This requires us to solve the constraint, $\partial_y\mathcal{A}_x=-\frac{1}{2}J$. By setting
\begin{equation}
    J(\mathbf{x})=\sum_I \lambda_I\delta^2(\mathbf{x}-\mathbf{x}_I) \label{J lambda}
\end{equation}
where $I$ is an index that represents different insertions of the field $\phi(\mathbf{x}_I)$ on the worldsheet and $\lambda_I$ is a functional dependent on 
 $\mathbf{x}_I$, one can obtain the solution as
\begin{equation}
    \mathcal{A}_x=\frac{1}{2}\sum_I \lambda_I\theta(y_I-y) \delta(x_I-x) \label{Ax}
\end{equation}
where $\theta(x)$ is a Heaviside step function.

On the boundary, the generating function takes the form 
\begin{equation}
    Z[J]=\widetilde{\mathcal{N}}\text{tr}\big(\mathcal{P}\big(\exp\big[-\frac{1}{2}\sum_I \lambda_I(\mathbf{x}_I)\big]\big)\big).
\end{equation}
Therefore, an expectation value of a product of $\phi$ on the boundary is
\begin{align}
      \langle \phi^{R_1}(x_1)\phi^{R_2}(x_2)\ldots\phi^{R_n}(x_n)\rangle_{\phi,\mathcal{A}}&=\frac{1}{Z[0]}\frac{\delta^n Z[J]}{\delta \lambda_{R_1}(x_1)\delta \lambda_{R_2}(x_2)\ldots\lambda_{R_n}(x_n)}\bigg\rvert_{J=0} \nonumber \\
      &=\bigg(-\frac{1}{2} \bigg)^n\mathcal{P} \, \text{tr}(T^{R_1}T^{R_2}\ldots T^{R_n}). \label{phi cor fn}
\end{align}

Although we are able to incorporate the boundary path-ordered generators into the string model, the object on the left-hand side of (\ref{phi cor fn}) is still not what we desire as  it is not gauge invariant which implies that the result depends on what gauge conditions we utilized. Accordingly, to make the observable gauge independent, we insert Wilson lines $\mathcal{W}_C$ defined as 
 \begin{equation}
    \mathcal{W}_C(\mathbf{x}_1,\mathbf{x}_2)=\mathcal{P}\exp({-\int_C \mathcal{A}\cdot d\omega})
\end{equation}
where $C$ is an arbitrary curve whose endpoints are $\mathbf{x}_1$ and $\mathbf{x}_2$. Therefore, one can modify the observables (\ref{phi cor fn}) to be 
\begin{equation}
    \text{tr}\bigg( \phi(\mathbf{x}_1) \mathcal{W}_{C_1}(\mathbf{x}_1,\mathbf{x}_2)\phi(\mathbf{x}_2)\mathcal{W}_{C_2}(\mathbf{x}_2,\mathbf{x}_3)\phi(\mathbf{x}_3)\cdots\phi(\mathbf{x}_n)\mathcal{W}_{C_n}(\mathbf{x}_n,\mathbf{x}_1)\bigg). \label{gauge inv obs}
\end{equation}
The appearance of trace and Wilson lines is to make the product gauge invariant.

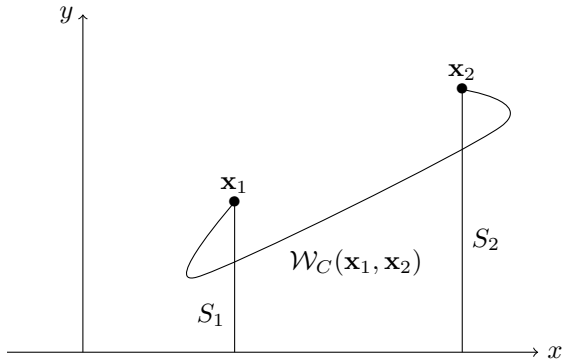
\begin{figure}
\centering
\begin{tikzpicture}
\draw[->] (-1,0)--(6,0);
\draw[->] (0,0)--(0,4.5);
\path (2,2) coordinate (x1);
\path (5,3.5) coordinate (x2);
\draw (2,0)--(x1);
\draw (5,0)--(x2);
\node at (x1) {\textbullet};
\node[above] at (x1) {$\mathbf{x}_1$};
\node[above] at (x2) {$\mathbf{x}_2$};
\node at (x2) {\textbullet};
\node[right] at (6,0) {$x$};
\node[left] at (0,4.5) {$y$};
\node[right] at (5,1.5) {$S_2$};
\node[left] at (2,0.5) {$S_1$};
 \draw  plot [smooth,tension=0.5] coordinates {(x1) (1.5,1) (5.5,3) (x2)};
\node[below] at (3.6,1.5) {$\mathcal{W}_C(\mathbf{x}_1,\mathbf{x}_2)$};
\end{tikzpicture}
    \caption{A Wilson line connecting from the point $\mathbf{x}_1$ to $\mathbf{x}_2$}
    \label{Wilson line}
\end{figure}

To evaluate (\ref{gauge inv obs}), let first consider just two insertions of $\phi$ at the points $\mathbf{x}_1$ and $\mathbf{x}_2$ with a Wilson line joining both points, i.e. 
\begin{equation*}
    \phi^{R_1}(\mathbf{x}_1)\mathcal{W}_C(\xi_1,\xi_2)\phi^{R_2}(\mathbf{x}_2),
\end{equation*}
in the axial gauge condition where we set $\mathcal{A}_y=0$ at each point. Using the expression (\ref{Ax}), the Wilson line $\mathcal{W}_C(\mathbf{x}_1,\mathbf{x}_2)$ is simply an ordered product of group elements when the contour $C$ intersects with the vertical lines $S_1$ and $S_2$ as illustrate in the figure \ref{Wilson line}. Surely, the value of the Wilson line depends on the path $C$ we are considering. For the given path shown in the figure, it reads $\exp{(-\lambda(\mathbf{x}_1))}\exp{(-\lambda(\mathbf{x}_2))}$. Note that the sign of the exponents rely on the orientation of the path $C$ when cutting through the vertical lines.

If the insertion points $\mathbf{x}_1$ and $\mathbf{x}_2$ are moved to the boundary, the Wilson line will reduce to one as no path can go below the inserted points. As a result, the boundary expectation value of (\ref{gauge inv obs}) is
\begin{equation}
    \bigg(-\frac{1}{2} \bigg)^n \bigg(\mathcal{P} \, \text{tr}(T^{R_1}T^{R_2}\ldots T^{R_n})\bigg) \text{tr}(T_{R_1}T_{R_2}\ldots T_{R_n}). \label{expect val gauge inv}
\end{equation}
Remember that the value (\ref{expect val gauge inv}) does not depend on which gauge choice and the contour paths we chose. 

Consequently, we propose a modification to the string contact interaction $S_I$ (\ref{string with contact action}) for incorporating the boundary path-ordered generators as
\begin{align}
    S_I^{\phi,\mathcal{A}}[X,\phi,\mathcal{A}]=2q^2\int_\Sigma d\Sigma_{\mu\nu}(\xi)d\Sigma^{\mu\nu}(\tilde{\xi}) \text{tr}\Big(& \phi(\xi)\langle\mathcal{W}_{C_1}(\xi,\tilde{\xi})\rangle_{C_1}\delta^4(X(\xi)-X(\tilde{\xi})) \nonumber \\
    &\times \phi(\tilde{\xi})\langle\mathcal{W}_{C_2}(\tilde{\xi},\xi)\rangle_{C_2}\Big) \label{new model}
\end{align}
This form of action enjoys gauge symmetry at the worldsheet level. To avoid picking a particular path of the Wilson lines, we average over all possible paths using the approach \cite{Curry:2017cnu}. The expectation value over all the paths $C$ is given by
\begin{equation}
    \langle \Omega \rangle_C=\mathcal{N}\int Dx \ \Omega e^{-S[x]} \label{curve avr}
\end{equation}
with 
\begin{equation}
     S[x]=\frac{1}{2}\int_0^\infty dt \ h_{rs}(x) \dot{x}^r \dot{x}^s
\end{equation}
where $h_{rs}$ is the induced metric along the path.

We then construct the functional integral to be
\begin{align}
   \Psi= \frac{1}{Z}\int D\phi D\mathcal{A} DX Dg &\exp\bigg(-S_{\text{P}}[X,g]-S_I^{\phi,\mathcal{A}}[X,\phi]-S_{\text{BF}}[\phi,\mathcal{A}]  \bigg) \nonumber \\
   &\times \text{tr}\big(\mathcal{P}\big(e^{-\oint_{\partial\Sigma} \mathcal{A}\cdot d\xi}\big)\big) \label{Psi}
\end{align}
with a suitable normalization factor $Z$. This functional integral is simply an expectation value of $\exp(-S_I^{\phi,\mathcal{A}})$ whose dynamics are described by the string action $S_P$ and the topological BF action $S_\text{BF}$ with a Wilson loop inserted along the boundary. 

In this paper, we aim to show that the expectation of the non-Abelian Wilson loop in gauge theory $\langle W[\partial\Sigma]\rangle$ is equivalent to the functional integral $\Psi$. However, we will not be able to give a full demonstration of the equality between the two models due to complications of potential divergences but rather show that the model (\ref{Psi}) contains the required ingredients to reproduce the non-Abelian gauge theory Wilson loop. A more complete proof needs to wait until the supersymmetric model is developed. 


\section{Evaluation of \texorpdfstring{$\Psi$}{Psi}: Boundary Contributions}\label{sec3}
In this section, we would like to evaluate the partition function $\Psi$ along the worldsheet boundary ignoring any contributions from the worldsheet interior. To evaluate $\Psi$, we first expand the exponential of the contact interaction $S_I^{\phi,\mathcal{A}}$ into the power series of $S_I^{\phi,\mathcal{A}}$ as
\begin{align}
   \Psi&= \frac{1}{Z}\int D(X,g,\phi, \mathcal{A})  e^{-(S_P[X,g]+S_\text{BF}[\phi,\mathcal{A}])} \bigg(1-S_I^{\phi,\mathcal{A}}+\frac{1}{2}(S_I^{\phi,\mathcal{A}})^2+\ldots \bigg) \nonumber \\
   &= \sum_{n=0}^\infty \frac{(-1)^n}{n!} \langle \big(S_I^{\phi,\mathcal{A}}\big)^n\rangle_{\Sigma,\phi\mathcal{A}}. \label{Psi expand}
\end{align}
Let first consider the first non-trivial term of (\ref{Psi expand}), i.e.
\begin{align}
        \langle S_I^{\phi,\mathcal{A}} \rangle_{\Sigma,\phi,\mathcal{A}}=\frac{1}{Z}\int D(X,g,\phi, \mathcal{A}) \ S_I^{\phi,\mathcal{A}} \ e^{-(S_P[X,g]+S_\text{BF}[\phi,\mathcal{A}])}\text{tr}\big(\mathcal{P}\big(e^{-q\oint_C \mathcal{A}\cdot d\xi}\big)\big). \label{avr SI}
\end{align}
To evaluate this, we rewrite  $S_I^{\phi,\mathcal{A}}$ (\ref{non-AB}) in terms of vertex operators $V_k^{\mu\nu}(\xi)$ as
\begin{align}
    S_I^{\phi,\mathcal{A}}=2q^2\int \frac{d^4k}{(2\pi)^4} \int_\Sigma \int_\Sigma d^2\xi  d^2\xi' \ \text{tr}\bigg(&\phi(\xi) \langle\mathcal{W}_{C_1}(\xi,\xi')\rangle_{C_1}V_k^{\mu\nu}(\xi)V_{-k \ \mu\nu}(\xi') \nonumber\\  
    &\times \phi(\xi')\langle\mathcal{W}_{C_2}(\xi',\xi)\rangle_{C_2}\bigg). \label{SI ver}
\end{align}
Similar to the Abelian calculation, the projection operator $\mathbb{P}_k$ in (\ref{proj op}) was used to write
\begin{align}
    \phi(\xi)\mathcal{W}_{C}(\xi,\xi')V_k^{\mu\nu}(\xi)\mathcal{F}(\xi)=&\phi(\xi)\mathcal{W}_{C}(\xi,\xi')\widetilde{V}_k^{\mu\nu}(\xi) \mathcal{F}(\xi) \nonumber \\*
    &+\partial_a( \phi \mathcal{W}_{C}(\xi,\xi')\mathcal{F}(\xi))\bigg(2i\epsilon^{ab} k^{[\mu}\partial_b \mathbb{P}_k(X)^{\nu]}\frac{e^{ik\cdot X}}{k^2} \bigg)  \nonumber \\* &-\partial_a\bigg(2i\phi \mathcal{W}_{C}(\xi,\xi')\epsilon^{ab} k^{[\mu}\partial_b \mathbb{P}_k(X)^{\nu]}\frac{e^{ik\cdot X}}{k^2} \mathcal{F}(\xi)\bigg)
\end{align}
with a generic function $\mathcal{F}(\xi)$. If we give $\mathcal{F}(\xi)=\phi(\xi')\mathcal{W}_{C'}(\xi',\xi)V_{-k \ \mu\nu}(\xi')$, one could find that
\begin{align}
    \phi(\xi)\mathcal{W}_{C}(\xi,\xi')&V_k^{\mu\nu}(\xi)\phi(\xi')\mathcal{W}_{C'}(\xi',\xi)V_{-k \ \mu\nu}(\xi')= \nonumber \\
    &\phi\mathcal{W}_{C}(\xi,\xi')\widetilde{V}_k^{\mu\nu}\phi'\mathcal{W}_{C'}(\xi',\xi)\widetilde{V}'_{-k \ \mu\nu}\nonumber \\
    &+\partial_a(\phi\mathcal{W}_{C}(\xi,\xi'))\mathcal{U}_k^{a \ \mu\nu} \phi'\mathcal{W}_{C'}(\xi',\xi)\widetilde{V}'_{-k \ \mu\nu}\nonumber \\
    &+\phi\mathcal{W}_{C}(\xi,\xi')\mathcal{U}_k^{a \ \mu\nu} \partial_a\Big( \phi'\mathcal{W}_{C'}(\xi',\xi)\widetilde{V}'_{-k \ \mu\nu}\Big)\nonumber \\
    &-\partial_a\Big(\phi\mathcal{W}_{C}(\xi,\xi')\mathcal{U}_k^{a \ \mu\nu} \phi'\mathcal{W}_{C'}(\xi',\xi)\widetilde{V}'_{-k \ \mu\nu}\Big) \nonumber \\
    &+\phi\mathcal{W}_{C}(\xi,\xi')\widetilde{V}_k^{\mu\nu}\partial'_a(\phi'\mathcal{W}_{C'}(\xi',\xi))\mathcal{U}'^a_{-k \ \mu\nu}\nonumber \\
    &+\partial'_a\Big(\phi\mathcal{W}_{C}(\xi,\xi')\widetilde{V}_k^{\mu\nu}\Big)\phi'\mathcal{W}_{C'}(\xi',\xi)\mathcal{U}'^a_{-k \ \mu\nu} \nonumber \\
    &-\partial'_a\Big(\phi\mathcal{W}_{C}(\xi,\xi')\widetilde{V}_k^{\mu\nu}\phi'\mathcal{W}_{C'}(\xi',\xi)\mathcal{U}'^a_{-k \ \mu\nu}\Big)\nonumber \\
    &+\partial_a(\phi\mathcal{W}_{C}(\xi,\xi'))\mathcal{U}_k^{a \ \mu\nu}\partial'_b(\phi'\mathcal{W}_{C'}(\xi',\xi))\mathcal{U}'^b_{-k \ \mu\nu} \nonumber \\
    &+\partial'_b\Big(\partial_a(\phi\mathcal{W}_{C}(\xi,\xi'))\mathcal{U}_k^{a \ \mu\nu}\Big)\phi'\mathcal{W}_{C'}(\xi',\xi)\mathcal{U}'^b_{-k \ \mu\nu}\nonumber \\
    &-\partial'_b\Big(\partial_a(\phi\mathcal{W}_{C}(\xi,\xi'))\mathcal{U}_k^{a \ \mu\nu}\phi'\mathcal{W}_{C'}(\xi',\xi)\mathcal{U}'^b_{-k \ \mu\nu}\Big) \nonumber \\
    &+\phi\mathcal{W}_{C}(\xi,\xi')\mathcal{U}_k^{a \ \mu\nu} \partial_a\Big(\partial'_b(\phi'\mathcal{W}_{C'}(\xi',\xi))\mathcal{U}'^b_{-k \ \mu\nu}\Big)\nonumber \\
    &+\partial'_b\Big(\phi\mathcal{W}_{C}(\xi,\xi')\mathcal{U}_k^{a \ \mu\nu}\Big)\partial_a\Big(\phi'\mathcal{W}_{C'}(\xi',\xi)\mathcal{U}'^b_{-k \ \mu\nu}\Big) \nonumber \\
    &-\partial'_b\Big(\phi\mathcal{W}_{C}(\xi,\xi')\mathcal{U}_k^{a \ \mu\nu}\partial_a\Big(\phi'\mathcal{W}_{C'}(\xi',\xi)\mathcal{U}'^b_{-k \ \mu\nu}\Big)\Big)\nonumber \\
    &-\partial_a\Big(\phi\mathcal{W}_{C}(\xi,\xi')\mathcal{U}_k^{a \ \mu\nu}\partial'_b(\phi'\mathcal{W}_{C'}(\xi',\xi))\mathcal{U}'^b_{-k \ \mu\nu} \Big)\nonumber \\
    &-\partial_a\Big(\partial'_b\Big(\phi\mathcal{W}_{C}(\xi,\xi')\mathcal{U}_k^{a \ \mu\nu}\Big)\phi'\mathcal{W}_{C'}(\xi',\xi)\mathcal{U}'^b_{-k \ \mu\nu} \Big)\nonumber \\
    &+\partial_a\partial'_b\Big(\phi\mathcal{W}_{C}(\xi,\xi')\mathcal{U}_k^{a \ \mu\nu}\phi'\mathcal{W}_{C'}(\xi',\xi))\mathcal{U}'^b_{-k \ \mu\nu} \Big) \label{vertex expand}
\end{align}
where 
\begin{equation}
    \mathcal{U}_k^{a \ \mu\nu}(\xi)=2i\epsilon^{ab} k^{[\mu}\partial_b \mathbb{P}_k(X)^{\nu]}\frac{e^{ik\cdot X}}{k^2}.
\end{equation}
Note that for the above expression we denote the primed and un-primed variables to be the objects determined on the point $\xi'$ and $\xi$ respectively. However, the terms from the third to the eighth lines of (\ref{vertex expand}) can be discarded as the contraction between $\mathcal{U}$ and $\widetilde{V}$ contains $k_\mu\widetilde{V}^{\mu\nu}$ which is zero. For simplification, we introduce the vertex operators as follows:
\begin{subequations}
\begin{align}
    \mathbb{V}^{\mu\nu}_k(\xi,\xi')&=\phi(\xi)\langle\mathcal{W}_{C}(\xi,\xi')\rangle_{C}\widetilde{V}_k^{\mu\nu}(\xi) \\
    \mathbb{B}^\mu_k(\tau,\xi)&= \phi(\tau) \langle\mathcal{W}_{C}(\tau,\xi)\rangle_{C}\mathbb{P}_k(\dot{\omega})^\mu(\tau)e^{ik\cdot \omega(\tau)} \\
    \mathbb{C}^\mu_k(\xi,\xi')&=\partial_a( \phi(\xi) \langle\mathcal{W}_{C}(\xi,\xi')\rangle_{C})\epsilon^{ab}\partial_b \mathbb{P}_k(X)^\mu(\xi) e^{ik\cdot X(\xi)}\\    
    \mathbb{D}^{a \mu}_k(\xi,\xi')&= \phi(\xi) \langle\mathcal{W}_{C}(\xi,\xi')\rangle_{C}\epsilon^{ab}\partial_b\mathbb{P}_k(X)^\mu(\xi)e^{ik\cdot X(\xi)}
\end{align}\label{vertex operators}
\end{subequations} 
where we use the parameter $\tau$ to parameterize the worldsheet boundary on which the field $X^\mu$ is $\omega^\mu$ and denote $\dot{\omega}$ as $\partial\omega/\partial\tau$. 

With these definitions and the expression (\ref{vertex expand}), we can rewrite the contact interaction (\ref{SI ver})  as
\begin{align}
   &S_I^{\phi,\mathcal{A}}=2q^2\int \frac{d^4k}{(2\pi)^4}\text{tr}\bigg[ 2\oint_{\partial\Sigma}\oint_{\partial\Sigma} d\tau d\tau' \frac{1}{k^2}\mathbb{B}^\mu_k(\tau,\tau') \mathbb{B}_{-k \ \mu}(\tau',\tau)\bigg] \nonumber \\
    &-4\int_\Sigma d^2\xi \oint_{\partial\Sigma} d\tau \ \frac{1}{k^2}\bigg(\mathbb{C}^\mu_k(\xi,\tau) \mathbb{B}_{-k \ \mu}(\tau,\xi)+\mathbb{D}^{a\mu}_{k}(\xi,\tau)\partial_a\mathbb{B}_{-k \ \mu}(\tau,\xi)\bigg) \nonumber \\
   &+\int_\Sigma\int_\Sigma d^2\xi  d^2\xi'\ \bigg(\mathbb{V}^{\mu\nu}_k(\xi,\xi') \mathbb{V}_{-k \ \mu\nu}(\xi',\xi)+\frac{2}{k^2} \mathbb{C}^\mu_k(\xi,\xi')\mathbb{C}_{-k\ \mu}(\xi',\xi)\bigg) \nonumber \\
    &+2\int_\Sigma\int_\Sigma d^2\xi  d^2\xi'\ \frac{1}{k^2} \bigg( 2\partial'_a\big(\mathbb{C}^\mu_k(\xi,\xi')\big)\mathbb{D}^{a \mu}_{-k}(\xi',\xi)+\partial'_a\mathbb{D}^{a \mu}_{k}(\xi,\xi')\partial_b\mathbb{D}^{b}_{-k \ \mu}(\xi',\xi)\bigg) \label{non-AB-contact}
\end{align}
Notice that the terms which involve the vertex operators $\mathbb{C}$ and $\mathbb{D}$ are not present in the Abelian model. When averaging (\ref{non-AB-contact}), it is likely that the bulk terms will get suppressed due to an appearance of $e^{\pm ikX}$ in the interior of the worldsheet. This is because the Wick theorem of the exponential
\begin{equation}
 e^{ik\cdot X} = :e^{ik\cdot X}:e^{-\alpha'\pi k^2 G(\xi,\xi)} \label{wick expo}
\end{equation}
vanishes in the worldsheet's interior at the tensionless limit. Remember that the Green's function at coincident points $G(\xi,\xi)$ is divergent in the worldsheet interior as discussed earlier. However, we do not expect that all the bulk terms get completely suppressed by this manner as there is possible that some contractions among the bulk terms might provide singularity to negate the suppression. This issue will be explored in the section five where a contraction from the worldsheet gauge fields provides a singularity necessary to reproduce a self-interaction in the Yang-Mills theory.

Considering only the boundary terms, the boundary expectation of the contact interaction is
\begin{align}
    4q^2&\int \frac{d^4k}{(2\pi)^4}\text{tr}\bigg[\oint_{\partial\Sigma}\oint_{\partial\Sigma} d\tau d\tau' \frac{1}{k^2} \mathbb{B}^\mu_k(\tau,\tau') \mathbb{B}_{-k \ \mu}(\tau',\tau) \bigg] \nonumber \\
    =4q^2&\int \frac{d^4k}{(2\pi)^4}\text{tr}\bigg[\oint_{\partial\Sigma}\oint_{\partial\Sigma} d\tau d\tau' \phi(\tau)\langle \mathcal{W}_{C_1}(\omega,\omega')\rangle_{C_1}\mathbb{P}_k(\dot{\omega})^\mu(\tau) \frac{e^{ik\cdot(\omega-\omega')}}{k^2} \nonumber \\
    & \hphantom{2cq^2\int \frac{d^4k}{(2\pi)^4}\text{tr}\bigg[\oint_{\partial\Sigma}\oint_{\partial\Sigma}} \times\phi(\tau')\langle \mathcal{W}_{C_2}(\omega',\omega)\rangle_{C_2}\mathbb{P}_{-k}(\dot{\omega}')_\mu(\tau')\bigg] \label{boundary SI}
\end{align}
where we give $\omega\equiv\omega(\tau)$ and $\omega'\equiv\omega(\tau')$. According to (\ref{expect val gauge inv}), one finds
\begin{equation}
    \bigg\langle \text{tr}\Big( \phi(\tau)\langle \mathcal{W}_{C_1}(\omega,\omega')\rangle_{C_1}\phi(\tau')\langle \mathcal{W}_{C_2}(\omega',\omega)\rangle_{C_2}\Big)\bigg\rangle_{\phi,\mathcal{A}}=\frac{1}{8}\mathcal{P}\text{tr}(T^R T_R).
\end{equation}
Consequently, the expectation of $S_I^{\phi,\mathcal{A}}$ takes the form
\begin{align}
   \langle S_I^{\phi,\mathcal{A}} \rangle_{\Sigma,\phi,\mathcal{A}}&= \mathcal{P}\text{tr}(T^R T_R) \ \frac{q^2}{2}\int \frac{d^4k}{(2\pi)^4} \oint_{\partial\Sigma}\oint_{\partial\Sigma} d\tau d\tau'\mathbb{P}_k(\dot{\omega})^\mu \frac{e^{ik\cdot(\omega-\omega')}}{k^2} \mathbb{P}_{-k}(\dot{\omega}')_\mu \nonumber \\
   &=\mathcal{P}\text{tr}(T^R T_R)\langle S_I \rangle_\Sigma
   \label{leading non-AB contact}
\end{align}
where $\langle S_I \rangle_\Sigma$ is the expectation of the Abelian contact interaction expressed in (\ref{SI3}).

The expression (\ref{leading non-AB contact}) is the exact prescription to reformulate the expectation value of the non-Abelian Wilson loop omitting self-interaction terms. This can be seen by evaluating the boundary contributions of the expectation value of $e^{-S_I^{\phi,\mathcal{A}}}$ which is
\begin{align}
    &\sum_{n=0}^\infty\frac{\langle (-S_I^{\phi,\mathcal{A}})^n  \rangle_{\phi,\mathcal{A},\Sigma}}{n!}=\mathcal{P} \text{tr} \sum_{n=0}^\infty \frac{q^{2n}}{2^n n!}   \nonumber \\
    &\times\prod_{i=1}^n\bigg( \int \frac{d^4 k_i}{(2\pi)^4} \oint \oint d\mathbb{P}_{k_i}(X)^\mu(\xi) d\mathbb{P}_{k_i}(X)_\mu(\xi') \frac{e^{ik_i\cdot (X(\xi)-X(\xi')}}{k_i^2}  T^{R_i}T^{R_i}\bigg). \label{avr exp SI wo int}
\end{align}
 At the order $q^{2n}$, the expression (\ref{avr exp SI wo int}) describes a Wilson loop with $n$ pairs of gauge propagators which freely propagate between the boundary. The expression of the non-Abelian Wilson loop is presented in Appendix \ref{non AB WL}. Remember that the calculation was evaluated on the worldsheet boundary ignoring the effect of bulk terms in the contact interaction on which potential divergences may arise. In the Abelian case, such terms are absent by introducing supersymmetry on the worldsheet. However, a similar generalization of the non-Abelian case has not yet been achieved.

\section{Effective BF Theory and Worldsheet Gauge Propagator}\label{sec4}
To verify if the proposed string model provides a valid description of the non-Abelian Yang-Mills theory or not, we need to find out whether
the model can reproduce self-interactions of the gauge fields. According to the previous section, it is obvious that no such terms appear when evaluating solely on the worldsheet boundary. We expect that such terms can be obtained from the worldsheet gauge fields $\mathcal{A}$ in the worldsheet interior. To demonstrate this, it requires the computation of a worldsheet gauge field propagator. This can be obtained by inspecting the generating functional 
\begin{align}
   Z_\text{BF}[\mathcal{J}]= \frac{1}{\text{Vol}}\int D\phi D\mathcal{A} &\exp\bigg(-S_{\text{BF}}[\phi,\mathcal{A}]-2\int_\Sigma d^2\xi \epsilon^{ij}\text{tr}\big(\mathcal{J}_i\mathcal{A}_j\big)  \bigg). \label{gen fn A}
\end{align}
We continue the calculation by integrating out the gauge field $\mathcal{A}$. To do this, we expand the fields $\phi$ and $\mathcal{A}_i$ in the Cartan-Weyl basis as
\begin{equation}
    \phi=\phi_aH^a \qquad \text{and} \qquad \mathcal{A}_i=\chi_{ia}H^a+a_{i\alpha} E^\alpha.
\end{equation}
where $H^a$ and $E^\alpha$ are the Cartan and Weyl generators respectively satisfying the following algebra:
\begin{align}
    [H^a,H^b]=0, \qquad [H^a,E^\alpha]=\alpha^{(a)} E^\alpha, 
    \nonumber \\
    \text{and} \qquad [E^\alpha,E^\beta]=    \begin{cases}
      N^{\alpha\beta}E^{\alpha+\beta} & \text{if} \ \alpha+\beta\in \Phi\\
      H^\alpha & \text{if} \ \alpha+\beta=0\\
    \end{cases} 
\end{align}
where $H^\alpha$ is defined as $H^\alpha=\alpha_{a}H^a$. The Cartan generators $H^a$ are diagonal traceless matrices in the adjoint representation. Note that these bases are $\xi$-dependent. The Cartan generators were chosen such that the field $\phi$ lies within their subalgebra at every point on the surface.

To relate Lie indices $A$ with the Cartan and Weyl indices $a$ and $\alpha$, we introduce unit vectors $\hat{H}^a_A$ and $\hat{E}^\alpha_A$ in Lie vector space which are defined as $\delta^a_A$ and $\delta^\alpha_A$ respectively. As a result, the inner products among the vectors are
\begin{equation}
    \hat{H}^a_A\hat{H}^{Ab}=\eta^{ab}, \qquad \hat{E}^\alpha_A\hat{E}^{A\beta}=\eta^{\alpha\beta},\qquad  \hat{H}^a_A\hat{E}^{A\alpha}=0
\end{equation}
and the completeness relation is
\begin{equation}
     \hat{H}^A_a\hat{H}^a_B+\hat{E}^A_\alpha\hat{E}^\alpha_B=\delta^{A}_B. \label{completeness}
\end{equation}
It is not hard to write the field $\phi$ and $\mathcal{A}_i$ in terms of the unit vectors as
\begin{equation}
     \phi^A=\phi^a\hat{H}^A_a \qquad \text{and} \qquad \mathcal{A}_i^A=\chi^a_i\hat{H}^A_a+a^\alpha_i \hat{E}^A_\alpha. \label{Lie comp}
\end{equation}

Using the relations (\ref{Lie comp}), one can find the topological BF action (\ref{BF}) taking the form 
\begin{align}
    S_{\text{BF}}[\phi,\mathcal{A}]=\int_{\Sigma} d^2\xi \ \epsilon^{ij}\Bigg(&i f^{ABC} \phi\indices{_C} a^\alpha_{i} a^\beta_{j} \hat{E}_{\alpha A}\hat{E}_{\beta B}  -2(\partial_i\phi_A)a^\alpha_j\hat{E}^A_\alpha \nonumber \\*
    &-2(\partial_i\phi_A)\chi^a_j\hat{H}^A_a +2\partial_i(\phi\indices{_A}\chi_j^a\hat{H}^A_a) \Bigg).  \label{topological field action3}
\end{align}
Notice that there is no contribution from diagonal components of $\mathcal{A}_i^A$ to the first term as the Cartan subalgebra is commutative. The last term only contributes to the boundary of the worldsheet. Consequently, this gives the generating function (\ref{gen fn A}) as
\begin{equation}
    Z_\text{BF}[\mathcal{J}]=\frac{1}{\text{Vol}'}\int D\phi\indices{_A} Da^\alpha_i  D\chi_{ja} \ \text{exp}(-S[\mathcal{J}]). \label{partition function with source}
\end{equation}
with
\begin{align}
    S[\mathcal{J}]=\int_\Sigma d^2\xi  \Big(&i f^{ABC} \phi\indices{_C} a^\alpha_{i} a^\beta_{j} \hat{E}_{\alpha A}\hat{E}_{\beta B}  -(2\partial_i\phi_A-\mathcal{J}_{iA})a^\alpha_j\hat{E}^A_\alpha \nonumber \\
    &-(2\partial_i\phi_A-2\phi_A\delta_{1i}\delta(\xi^1)-\mathcal{J}_{iA})\chi_{ja}\hat{H}^{Aa}\Big) \epsilon^{ij}. 
\end{align}
To obtain the above expression, we consider the case that the worldsheet $\Sigma$ has the
topology of a disk, accordingly, one can parametrize the worldsheet coordinates such that $\xi^1$ is zero at the boundary. In this way, the coordinates $\xi^1$ and $\xi^2$ can be seen as radial and angular-like coordinates.

The path integration of the second line gives a constraint on the theory in the form of the Dirac delta function
\begin{equation}
    \prod_{a=1}^{m} \prod_{i=1}^2\delta(\text{tr}((2\partial_i\phi-2\phi\delta_{1i}\delta(\xi^1)-\mathcal{J}_i)H^a)) 
\end{equation}
with $m$ the number of Cartan generators. Without the source $\mathcal{J}$, the constraint implies that the square of the field $\phi$ is constant throughout the worldsheet interior with a jump at the boundary, i.e.
\begin{equation}
    \lvert\phi\rvert^2\bigg\rvert_{\text{interior}}=3\rvert\phi\rvert^2\bigg\rvert_{\text{boundary}}.
\end{equation}
We then proceed the calculation by changing the worldsheet coordinates $(\xi^1,\xi^2)$ into the complex coordinates $(z,\bar{z})$ where $z=\xi^1+i\xi^2$ and $\bar{z}=\xi^1-i\xi^2$. In these new coordinates, the field $a^\alpha_i$ becomes  complex fields $b^\alpha$ where
\begin{equation}
    b^\alpha=\frac{1}{2}(a^\alpha_1-ia^\alpha_2) \qquad \text{and} \qquad \bar{b}^\alpha=\frac{1}{2}(a^\alpha_1+ia^\alpha_2). \label{complex b field}
\end{equation}
The partition function now resembles a Gaussian path integral with respect to the  fields $b$ and $\bar{b}$ which is 

\begin{align}
    Z_\text{BF}[\mathcal{J}]=\frac{\mathcal{N}}{\text{Vol}}\int D\phi\indices{_A} &Db^\alpha D\bar{b}^\alpha \exp{-S[\phi,b,\bar{b},\mathcal{J}]} \nonumber \\*
    &\times\prod_{i=1}^2\delta^{(m)}(\text{tr}((2\partial_i\phi-2\phi\delta_{1i}\delta(\xi^1)-\mathcal{J}_i)\hat\phi)). \label{partition function with souce complex}
\end{align}
where
\begin{align}
    S[\phi,b,\bar{b},\mathcal{J}]=\int_\Sigma d^2z \Big( &2if^{ABC}\phi\indices{_C} b^\alpha\bar{b}^\beta \hat{E}_{\alpha A}\hat{E}_{\beta B}\nonumber \\ &-((2\partial\phi_A-\mathcal{J}_{A})\bar{b}^\alpha-(2\bar{\partial}\phi_A-\bar{\mathcal{J}}_A)b^\alpha)\hat{E}^A_\alpha \Big).
\end{align}
$\hat{\phi}^A$ is a unit vector in Lie indices space of the field $\phi$ whose value is $\hat{H}^A_a H^a$. We can then use the Gaussian integration formula to integrate out the complex field $b$,
\begin{align}
    \int Db D\bar{b} \ e^{-\int d^2z(-\bar{b}^\alpha M_{\alpha\beta} b^\beta+\bar{J}_\alpha b^\alpha+J_\alpha \bar{b}^\alpha)}=\mathcal{N}_0\frac{e^{-\int d^2z (\bar{J}_\alpha (M^{-1})^{\alpha\beta} J_\beta}}{\prod\limits_{\forall \xi}\text{det}(M)}. \label{gaussian}
\end{align}
To apply (\ref{gaussian}) into (\ref{partition function with souce complex}), one can use a general expression for an inverse matrix $(\widetilde{M}^{-1})\indices{^\alpha_\beta}$ which is 
\begin{equation}
    (\widetilde{M}^{-1})\indices{^\alpha_\beta}=\frac{\text{adj}(\widetilde{M})\indices{^\alpha_\beta}}{\text{det}(\widetilde{M})} \label{inverse}
\end{equation}
where
\begin{align}
    \text{adj}(\widetilde{M})\indices{^\alpha_\beta}=& \ \delta^{\alpha j_2\ldots j_n}_{\beta i_2\ldots i_n} \  \widetilde{M}\indices{^{i_2}_{j_2}}\widetilde{M}\indices{^{i_3}_{j_3}}\ldots \widetilde{M}\indices{^{i_n}_{j_n}}, \nonumber \\ 
    \text{det}(\widetilde{M})=& \ \delta^{j_1j_2\ldots j_n}_{i_1 i_2\ldots i_n}\ \widetilde{M}\indices{^{i_1}_{j_1}}\widetilde{M}\indices{^{i_2}_{j_2}}\ldots \widetilde{M}\indices{^{i_n}_{j_n}}. \label{adj det}
\end{align}
$\delta^{j_1j_2\ldots j_n}_{i_1 i_2\ldots i_n}$ is a generalized Kronecker delta that is related to an anti-symmetrization of ordinary Kronecker deltas as
\begin{equation}
    \delta^{j_1j_2\ldots j_n}_{i_1 i_2\ldots i_n}=n!\delta^{j_1}_{[i_1}\delta^{j_2}_{i_2}\ldots\delta^{j_n}_{i_n]}. \label{generalised Kronecker delta}
\end{equation}
The integer $n$ is the number of Weyl generators. In the case of $SU(N)$, $n$ is equal to $N^2-N$. This yields
\begin{equation}
    Z[\mathcal{J}]=\frac{\mathcal{N}}{\text{Vol}}\int D\phi \ \prod_{i=1}^2\delta^{(m)}(\text{tr}((2\partial_i\phi-2\phi\delta_{1i}\delta(\xi^1)-\mathcal{J}_i)\hat\phi))\frac{\text{exp}\Big(-S_\text{eff}(\phi,\mathcal{J})\Big)}{\prod\limits_{\forall \xi}\text{det}(\widetilde{M})} \label{partition function with souce complex2}
\end{equation}
with
\begin{equation}
S_\text{eff}(\phi,\mathcal{J})=\int d^2z \ \frac{i}{2} \frac{1}{\text{det}(\widetilde{M})}(2\partial\phi\indices{_A}-\mathcal{J}_A) \Theta^{AB} (2\bar{\partial}\phi\indices{_B}-\bar{\mathcal{J}}_B)  \label{complex effective with source}
\end{equation}
where the matrix component $\widetilde{M}_{\alpha\beta}$ is given by
\begin{equation}
    \widetilde{M}_{\alpha\beta}=f^{ABC}\phi\indices{_B}\hat{E}_{\alpha A}\hat{E}_{\beta C} \label{M matrix}
\end{equation}
and the matrix $\Theta^{AB}$ relates to the adjugate matrix adj($\widetilde{M}$) by the relation
\begin{equation}
    \Theta^{AB}=\hat{E}^{A}_\alpha(\text{adj}(\widetilde{M})\indices{^\alpha_\beta})\hat{E}^{\beta B}. \label{Theta matrix}
\end{equation}
Turning back to the $(\xi^1,\xi^2)$ coordinates, the effective action takes the form
\begin{equation}
S_\text{eff}(\phi,\mathcal{J})=\int d^2 \xi \ \frac{i}{4} \frac{1}{\text{det}(\widetilde{M})}(2\partial_i\phi\indices{_A}-\mathcal{J}_{iA}) \Theta^{AB} (2\partial_j\phi\indices{_B}-\mathcal{J}_{jB})\epsilon^{ij}.  \label{effective with source}
\end{equation}
When $\mathcal{J}=0$, the expressions (\ref{complex effective with source}) and (\ref{effective with source}) are the effective actions of the BF theory provided earlier in \cite{Srisangyingcharoen:2021ndd} in which the general expression for the effective Lagrangian was presented in a diagrammatic representation.

One can include a boundary Wilson loop $\mathcal{W}_{\partial\Sigma}$ into the generating functional (\ref{gen fn A}) by introducing auxiliary either commuting or anti-commuting fields onto a worldline to generate the path-ordering of the Lie algebra generators \cite{Corradini:2016czo, Edwards:2016acz, Bastianelli:2015iba}. According to \cite{Samuel:1978iy,Broda:1995wv}, a trace together with a path-ordering operator can be replaced by a functional integral over the Grassmannian field $\psi$ as
\begin{equation}
    W_{\partial\Sigma}= \int D\psi^\dagger D\psi \exp \Big(\int d\tau \psi^\dagger \dot{\psi} -\mathcal{A}_{iR} \dot{\xi}^i \psi^\dagger T^R \psi \Big)
\end{equation}
where the loop $\partial\Sigma$ is now parameterized by $\tau$. Therefore, a suitable choice of the source term $\mathcal{J}$ to regain the notion of the Wilson loop $\mathcal{W}_{\partial\Sigma}$ is
\begin{equation}
    \widetilde{\mathcal{J}}\indices{^A_i}(\xi)=-\oint_{\partial\Sigma} \psi^\dagger(\tilde{\xi}) T^A \psi(\tilde{\xi}) \delta^{(2)}(\xi-\tilde{\xi})\epsilon_{ij} d\tilde{\xi}^j. \label{wilson loop source}
\end{equation}
Hence, to align with our model (\ref{Psi}), we modify the generating functional (\ref{gen fn A}) to be
\begin{IEEEeqnarray}{rCll}
   Z_\text{BF}[\widetilde{\mathcal{J}}]&=& \frac{1}{\text{Vol}}\int D\psi^\dagger D\psi D\phi &D\mathcal{A} \exp\bigg(-S_{\text{BF}}[\phi,\mathcal{A}]\nonumber \\
   && \qquad&-2\int_\Sigma d^2\xi \epsilon^{ij}\text{tr}\big(\widetilde{\mathcal{J}}_i\mathcal{A}_j\big) 
   +\int d\tau \psi^\dagger \dot{\psi}  \bigg) \nonumber \\
   &=&\IEEEeqnarraymulticol{2}{l} {\frac{1}{\text{Vol}}\int D(\psi^\dagger,\psi,\phi) \prod_{i=1}^2\delta^{(m)}(\text{tr}((2\partial_i\phi-2\phi\delta_{1i}\delta(\xi^1)-\widetilde{\mathcal{J}}_i)\hat\phi))}\nonumber \\  &&\qquad &\times\frac{\text{exp}\Big(-S_\text{eff}(\phi,\widetilde{\mathcal{J}})+\int d\tau \psi^\dagger \dot{\psi} \Big)  }{\prod\limits_{\forall \xi}\text{det}(\widetilde{M})}\label{gen fn A wil WL}
\end{IEEEeqnarray}
where $ D(\psi^\dagger,\psi,\phi)=D\psi^\dagger D\psi D\phi$. We can then obtain a two-point function of the fields $\mathcal{A}$ by
\begin{equation}
    \langle \mathcal{A}_i^A(\xi_1) \mathcal{A}_j^B(\xi_2) \rangle_\mathcal{A} =\frac{1}{Z_\text{BF}[\widetilde{\mathcal{J}}]} \epsilon_{im}\epsilon_{jn} \frac{\delta^2 Z_\text{BF}[\mathcal{J}]}{\delta \mathcal{J}_{Am}(\xi_1)\delta \mathcal{J}_{Bn}(\xi_2)}\bigg\rvert_{\mathcal{J}=\widetilde{\mathcal{J}}}. \label{A prop}
\end{equation}
It turns out that only the gauge propagator in the Weyl directions survives because when taking the functional derivative of $Z_\text{BF}[\mathcal{J}]$ with respect to $\mathcal{J}_{am}$ with a Cartan index $a$, it would appear terms containing $\Theta^{aB} \mathcal{J}_B$ in the integrand which is equal to zero due to the commutative property of the Cartan generators. Remind that a derivative of the Dirac delta functionals with respect to $\mathcal{J}$ gives the path integral zero as they contain only the source terms in the Cartan directions.

For this reason, the double functional derivative of (\ref{gen fn A wil WL}) takes the form
\begin{align}
   \frac{\epsilon_{im}\epsilon_{jn}}{Z_\text{BF}[\widetilde{\mathcal{J}}]}\int &D(\psi^\dagger,\psi,\phi)e^{\int d\tau \psi^\dagger \dot{\psi}}\prod_{i=1}^2  \frac{\delta^{(m)}(\mathcal{G}_i(\mathcal{J}))}{\prod\limits_{\forall \xi}\text{det}(\widetilde{M})} \nonumber \\
   &\times\bigg[\frac{\delta^2}{\delta \mathcal{J}_{Am}(\xi_1)\delta \mathcal{J}_{Bn}(\xi_2)}\text{exp} 
   \Big(-S_\text{eff}(\phi,\mathcal{J})\Big)\bigg]\bigg\rvert_{\mathcal{J}=\widetilde{\mathcal{J}}}
\end{align}
where $\mathcal{G}_i(\widetilde{\mathcal{J}})$ is the constraint defined as 
\begin{equation}
    \text{tr}((2\partial_i\phi-2\phi\delta_{1i}\delta(\xi^1)-\widetilde{\mathcal{J}}_i)\hat\phi). \label{boundary constraint}
\end{equation}
Taking the term in the square bracket into consideration, we obtain
\begin{align}
    \frac{\delta^2 \ e^{-S_{\text{eff}}(\phi,\mathcal{J})}}{\delta \mathcal{J}_{Am}(\xi_1)\delta \mathcal{J}_{Bn}(\xi_2)}=&\frac{i}{2}\frac{1}{\text{det}\widetilde{M}}\Theta^{AB}\epsilon^{mn}\delta^{(2)}(\xi_1-\xi_2)e^{-S_{\text{eff}}(\phi,\mathcal{J})}\nonumber \\*
    &-\frac{1}{4}\Big[\frac{1}{\text{det}\widetilde{M}}\Theta^{AC}(2\partial_r\phi_C-J_{rC})\Big](\xi_1) \nonumber \\*&\times\Big[\frac{1}{\text{det}\widetilde{M}}\Theta^{BD}(2\partial_s\phi_D-J_{sD})\Big](\xi_2)\epsilon^{mr}\epsilon^{ns}e^{-S_{\text{eff}}(\phi,\mathcal{J})} \label{double der}
\end{align}

If we consider the gauge propagator in the interior of the worldsheet, the two-point function becomes 
\begin{align}
    \langle \mathcal{A}_i^A(\xi_1) \mathcal{A}_j^B(\xi_2) \rangle_\mathcal{A}=&\frac{-i}{2}\frac{1}{\text{det}(\widetilde{\mathcal{M}})}\Theta^{AB}\epsilon_{ij}\delta^2(\xi_1-\xi_2)\nonumber \\
    &-\bigg[\frac{1}{\text{det}\widetilde{M}}\partial_i\phi_C\Theta^{AC}\bigg](\xi_1)\bigg[\frac{1}{\text{det}\widetilde{M}}\partial_j\phi_D\Theta^{BD}\bigg](\xi_2).\label{A prop2}
\end{align}
To obtain the above relation, (\ref{double der}) and the fact that the source term inside the worldsheet is zero were used. For the case of $SU(2)$, one can find that the matrix $\Theta^{AB}=-\epsilon^{ACB}\phi_C$ and the matrix determinant is $2\rvert\phi\rvert^2$. More detail can be found in \cite{Srisangyingcharoen:2021ndd}. Thus, (\ref{A prop2}) becomes
\begin{align}
    \langle \mathcal{A}_i^A(\xi_1) \mathcal{A}_j^B(\xi_2) \rangle_\mathcal{A}^{SU(2)}=&\frac{i}{4}\frac{1}{\rvert\phi\rvert^2}\epsilon^{ACB}\phi_C\epsilon_{ij}\delta^2(\xi_1-\xi_2) \nonumber \\
    &-\frac{1}{4}\bigg[\partial_i\hat{\phi}_D\epsilon^{ACD}\hat{\phi}_C\bigg](\xi_1)\bigg[\partial_j\hat{\phi}_F\epsilon^{BEF}\hat{\phi}_E\bigg](\xi_2). \label{A prop su2}
\end{align}
In this case, the effective action for SU(2) BF theory is
\begin{equation}
   S_{\text{eff}}^{SU(2)}(\phi)= \int_\Sigma d^2\xi \frac{i}{2\rvert\phi\rvert^2} \partial_i\phi\indices{_A}\partial_j\phi\indices{_B} \epsilon^{ij}\phi_C\epsilon\indices{^A^B^C}. \label{SU2 eff2}
\end{equation}
Note that to get the second line in (\ref{A prop su2}), the constraint $\mathcal{G}_i(0)=0$ was implemented which implies that $\rvert\phi\rvert$ is constant throughout the worldsheet $\Sigma$.

\section{Evaluation of \texorpdfstring{$\Psi$}{Psi}: Three-point Self-interactions}\label{sec5}
In this section, we would like to determine whether or not our string model can re-create self-interaction vertices of the gauge theory. To demonstrate this, we provide a perturbative computation of $\Psi$ up to a few orders of $q$ where the effects of self-interactions can be observed. 

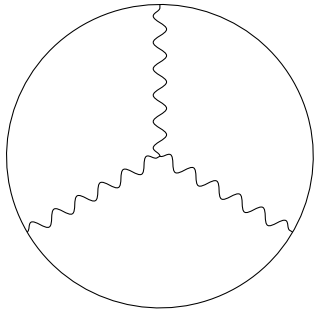
\begin{figure}
\centering
\begin{tikzpicture}
\def\h{3.5}
\path (0,0) coordinate(A);
\path (0,0)++(0:\h) coordinate(B);
\path (0,0)++(60:\h) coordinate(C);
\path (0,0) ++(30:\h/1.732) coordinate(D);
\draw (A) arc(-150:210:\h/1.732);
\draw [snake=snake](A)--(D)--(B);
\draw [snake=snake](D)--(C);
\end{tikzpicture}
    \caption{The simplest Wilson loop with the three gauge boson vertex}
    \label{3 vertex WL}
\end{figure}

For simplicity, we content ourselves to work in the simplest case of the non-Abelian model, i.e. $SU(2)$. Since a contribution of the three gauge-boson vertex in the Wilson loop is first observed at $\mathcal{O}(q^4)$, we expect that the expectation of $(S_I^{\phi,\mathcal{A}})^2$ contains the expression for the Wilson loop with three-point interaction as illustrated in the figure \ref{3 vertex WL}. The question is which terms can generate such a structure. As mentioned earlier that it is highly probable that the bulk terms will get suppressed in the tensionless limit unless there exist any singular terms which potentially spoil the suppression. According to (\ref{A prop2}), we got a hint that Wick contraction of the worldsheet gauge fields $\mathcal{A}$ generates singularity at a coincident point. This could result in forming a new vertex operator at the coincident point. By this argument, we speculate that the three-point vertex interaction to appear in the expectation of
\begin{align}
   &\frac{1}{2}(S_I^{\phi,\mathcal{A}})^2\ni 32q^4\int \frac{d^4k}{(2\pi)^4} \frac{1}{k^2}\int \frac{d^4k'}{(2\pi)^4} \frac{1}{k'^2} \nonumber \\
   &\times\text{tr}\bigg[\int_\Sigma d^2\xi \oint_{\partial\Sigma} d\tau \bigg( \mathbb{B}_{k \ \mu}(\tau,\xi)\mathbb{C}^\mu_{-k }(\xi,\tau)+\mathbb{D}^{a\mu}_{k}(\xi,\tau)\partial_a\mathbb{B}_{-k \ \mu}(\tau,\xi)\bigg) \bigg] \nonumber \\
   & \times\text{tr}\bigg[\int_\Sigma d^2\xi' \oint_{\partial\Sigma} d\tau' \bigg( \mathbb{B}_{k' \ \nu}(\tau',\xi')\mathbb{C}^{\nu}_{-k'}(\xi',\tau')+\mathbb{D}^{a\nu}_{k'}(\xi',\tau')\partial'_a\mathbb{B}_{-k' \ \nu}(\tau',\xi')\bigg)\bigg] \label{3-pt ver SI}
\end{align}
where the vertex operators were defined in (\ref{vertex operators}). To evaluate the right-hand side of (\ref{3-pt ver SI}), we begin by considering the products $ \mathbb{C}^\mu_{-k }(\xi,\tau)\mathbb{C}^{\nu}_{-k'}(\xi',\tau')$, $ \mathbb{C}^\mu_{-k }(\xi,\tau)\partial'_a\mathbb{B}^\nu_{-k'}(\tau',\xi')$ and $\partial_a\mathbb{B}^\mu_{-k}(\tau,\xi)\partial'_b\mathbb{B}^\nu_{-k'}(\tau',\xi')$. One can show that these products contain the following terms:
\begin{subequations}
\begin{IEEEeqnarray}{rll}
 \mathbb{C}^\mu_{-k }(\xi,\tau)\mathbb{C}^{\nu}_{-k'}&(\xi',\tau') &\ni \  \phi(\xi)\mathcal{A}_a(\xi)\langle\mathcal{W}_{C}(\xi,\tau)\rangle_{C}\epsilon^{ab}\partial_b \mathbb{P}_{-k}(X)^\mu(\xi) e^{-ik\cdot X(\xi)} \nonumber \\
  &\IEEEeqnarraymulticol{2}{l}{\times \phi(\xi')\mathcal{A}_c(\xi')\langle\mathcal{W}_{C'}(\xi',\tau')\rangle_{C'}\epsilon^{cd}\partial'_d \mathbb{P}_{-k'}(X)^\nu(\xi') e^{-ik'\cdot X(\xi')}}\label{contract CC} \\
 \IEEEeqnarraymulticol{2}{l}{\mathbb{C}^\mu_{-k }(\xi,\tau)\partial'_a\mathbb{B}^\nu_{-k'}(\tau',\xi') } &\ni \  \phi(\xi)\mathcal{A}_b(\xi)\langle\mathcal{W}_{C}(\xi,\tau)\rangle_{C}\epsilon^{bc}\partial_c \mathbb{P}_{-k}(X)^\mu(\xi) e^{-ik\cdot X(\xi)} \nonumber\\
 &  \IEEEeqnarraymulticol{2}{l}{\times \phi(\tau')\langle\mathcal{W}_{C'}(\tau',\xi')\rangle_{C'}\mathcal{A}_a(\xi')\mathbb{P}_{-k'}(\dot{\omega})^\nu(\tau')e^{-ik'\cdot\omega(\tau')}}\label{contract CB} \\
  \IEEEeqnarraymulticol{2}{l}{\partial_a\mathbb{B}^\mu_{-k}(\tau,\xi)\partial'_b\mathbb{B}^\nu_{-k'}(\tau',\xi')} & \ni \ \phi(\tau)\langle\mathcal{W}_{C}(\tau,\xi)\rangle_{C}\mathcal{A}_a(\xi)\mathbb{P}_{-k}(\dot{\omega})^\mu(\tau)e^{-ik\cdot\omega(\tau)} \nonumber \\
 & \IEEEeqnarraymulticol{2}{l}{ \times \phi(\tau')\langle\mathcal{W}_{C'}(\tau',\xi')\rangle_{C'}\mathcal{A}_b(\xi')\mathbb{P}_{-k'}(\dot{\omega})^\nu(\tau')e^{-ik'\cdot\omega(\tau')}} \label{contract BB}
\end{IEEEeqnarray}
\end{subequations}
whose values are singular at the point $\xi=\xi'$ through the Dirac delta function when Wick contracted using (\ref{A prop su2}). We can then apply Wick's theorem for the worldsheet gauge field $\mathcal{A}$ to determine the singular behavior of the above products. This turns (\ref{contract CC}), (\ref{contract CB}) and (\ref{contract BB}) to be
\begin{subequations}
\begin{align}
        \frac{i}{4}&\delta^2(\xi-\xi')\phi(\xi)T_A\langle\mathcal{W}_{C}(\xi,\tau)\rangle_{C}\phi(\xi')T_B\langle\mathcal{W}_{C'}(\xi',\tau')\rangle_{C'} \nonumber \\
    &\times\epsilon^{ACB}\frac{\phi_C}{\rvert\phi\rvert^2}(\xi)\epsilon^{ab}\partial_a \mathbb{P}_{-k}(X)^\mu(\xi)\partial'_b \mathbb{P}_{-k'}(X)^\nu(\xi')e^{-i(k'\cdot X(\xi')+k\cdot X(\xi))} \label{contract CC2}, \\
    \frac{i}{4}&\delta^2(\xi-\xi')\phi(\xi)T_A\langle\mathcal{W}_{C}(\xi,\tau)\rangle_{C}\phi(\xi')\langle\mathcal{W}_{C'}(\xi',\tau')\rangle_{C'}T_B \nonumber \\
     &\times\epsilon^{ACB}\frac{\phi_C}{\rvert\phi\rvert^2}(\xi)\partial_a \mathbb{P}_{-k}(X)^\mu(\xi) \mathbb{P}_{-k'}(\dot{\omega})^\nu(\tau')e^{-i(k'\cdot \omega(\tau')+k\cdot X(\xi))} \label{contract CB2}, \\
     \frac{i}{4}&\delta^2(\xi-\xi')\phi(\tau)\langle\mathcal{W}_{C}(\tau,\xi)\rangle_{C}T_A\phi(\tau')\langle\mathcal{W}_{C'}(\tau',\xi')\rangle_{C'}T_B \nonumber \\
     &\times\epsilon^{ACB}\frac{\phi_C}{\rvert\phi\rvert^2}(\xi)\epsilon_{ab} \mathbb{P}_{-k}(\dot{\omega})^\mu(\tau) \mathbb{P}_{-k'}(\dot{\omega})^\nu(\tau')e^{-i(k'\cdot \omega(\tau')+k\cdot \omega(\tau))} \label{contract BB2}
\end{align}
\end{subequations}
respectively. The above expressions are merely the singular parts among other non-singular terms resulting from the Wick's theorem. Note that there are still gauge fields left uncontracted in the Wilson lines which we will cope with them later. In addition, we have omitted the contribution of the second line of (\ref{A prop su2}) as it will be suppressed by the matter fields $X^\mu$ when functional averaged in the tensionless limit. Consequently, to obtain the expression for the right-hand side of (\ref{3-pt ver SI}), we need to evaluate the following integrands which are
\begin{subequations}
\begin{equation}
    \text{tr}\bigg[\mathbb{B}_{k \ \mu}(\tau,\xi)\mathbb{C}^\mu_{-k }(\xi,\tau)\bigg]\text{tr}\bigg[\mathbb{B}_{k' \ \nu}(\tau',\xi')\mathbb{C}^{\nu}_{-k'}(\xi',\tau')\bigg], \label{1st integrand} 
\end{equation}
\begin{equation}
        \text{tr}\bigg[\mathbb{B}_{k \ \mu}(\tau,\xi)\mathbb{C}^\mu_{-k }(\xi,\tau)\bigg]\text{tr}\bigg[\mathbb{D}^{a\nu}_{k'}(\xi',\tau')\partial'_a\mathbb{B}_{-k' \ \nu}(\tau',\xi')\bigg], \label{2nd integrand}
\end{equation}
and
\begin{equation}
        \text{tr}\bigg[\mathbb{D}^{a\mu}_{k}(\xi,\tau)\partial_a\mathbb{B}_{-k \ \mu}(\tau,\xi)\bigg]\text{tr}\bigg[\mathbb{D}^{b\nu}_{k'}(\xi',\tau')\partial'_b\mathbb{B}_{-k' \ \nu}(\tau',\xi')\bigg]. \label{3rd integrand}
\end{equation}
\end{subequations}

Substitute (\ref{contract CC2}) into the first integrand, one obtains 
\begin{align}
   & \frac{i}{4}\delta^2(\xi-\xi') \epsilon^{ACB}\frac{\phi_C}{\rvert\phi\rvert^2}(\xi)\epsilon^{ab}\bigg(\partial_a \mathbb{P}_{-k}(X)^\mu\partial'_b \mathbb{P}_{-k'}(X')^\nu\bigg)e^{-i(k\cdot X+k'\cdot X')}  \nonumber \\
   &\times \text{tr}\bigg[\mathbb{B}_{k \ \mu}(\tau,\xi) \phi(\xi)T_A\langle\mathcal{W}_{C}(\xi,\tau)\rangle_{C}\bigg]\text{tr}\bigg[\mathbb{B}_{k' \ \nu}(\tau',\xi)\phi(\xi')T_B\langle\mathcal{W}_{C'}(\xi,\tau')\rangle_{C'} \bigg] \nonumber \\
   =& \frac{i}{4}\delta^2(\xi-\xi') \epsilon^{ACB}\frac{\phi_C}{\rvert\phi\rvert^2}(\xi)\epsilon^{ab}\partial_a X^\mu(\xi)\partial'_b X'^\nu(\xi') \mathbb{P}_k(\dot{\omega})_\mu(\tau)\mathbb{P}_{k'}(\dot{\omega})_\nu(\tau')  \nonumber \\*
   &\times \text{tr}\bigg[\phi(\tau)\langle\mathcal{W}_C(\tau,\xi)\rangle_C \phi(\xi)T_A\langle\mathcal{W}_{C'}(\xi,\tau)\rangle_{C'}\bigg]e^{-i(k\cdot (X-\omega)+k'\cdot (X'-\omega'))} \nonumber \\*
   &\times\text{tr}\bigg[\phi(\tau')\langle\mathcal{W}_{\tilde{C}}(\tau',\xi')\rangle_{\tilde{C}} \phi(\xi')T_B\langle\mathcal{W}_{\tilde{C}'}(\xi',\tau')\rangle_{\tilde{C}'} \bigg]. \label{BCCB}
\end{align}
To obtain the last expression, we used the fact that $k\cdot d\mathbb{P}_k(X)=0$, hence $\partial_a\mathbb{P}_{-k}(X)^\mu \mathbb{B}_{k \ \mu}=\partial_a X^\mu \mathbb{B}_{k \ \mu}$.

Similarly, one can find the two remaining integrands, i.e. (\ref{2nd integrand}) and (\ref{3rd integrand}), as 
\begin{align}
& \frac{i}{4}\delta^2(\xi-\xi') \epsilon^{ACB}\frac{\phi_C}{\rvert\phi\rvert^2}(\xi)\epsilon^{ab}\partial_a X^\mu(\xi)\partial'_b X'^\nu(\xi')   \mathbb{P}_k(\dot{\omega})_\mu(\tau)\mathbb{P}_{-k'}(\dot{\omega})_\nu(\tau')  \nonumber \\*
    &\times \text{tr}\bigg[\phi(\tau)\langle\mathcal{W}_C(\tau,\xi)\rangle_C \phi(\xi)T_A\langle\mathcal{W}_{C'}(\xi,\tau)\rangle_{C'}\bigg]e^{-i(k\cdot (X-\omega)-k'\cdot (X'-\omega'))} \nonumber \\*
   &\times\text{tr}\bigg[\phi(\xi')\langle\mathcal{W}_{\tilde{C}}(\xi',\tau')\rangle_{\tilde{C}} \phi(\tau')\langle\mathcal{W}_{\tilde{C}'}(\tau',\xi')\rangle_{\tilde{C}'}T_B \bigg]\label{BCDB}
\end{align}
and
\begin{align}
& \frac{i}{4}\delta^2(\xi-\xi') \epsilon^{ACB}\frac{\phi_C}{\rvert\phi\rvert^2}(\xi)\epsilon^{ab}\partial_a X^\mu(\xi)\partial'_b X'^\nu(\xi')   \mathbb{P}_{-k}(\dot{\omega})_\mu(\tau)\mathbb{P}_{-k'}(\dot{\omega})_\nu(\tau')  \nonumber \\*
    &\times \text{tr}\bigg[\phi(\xi)\langle\mathcal{W}_C(\xi,\tau)\rangle_C \phi(\tau)\langle\mathcal{W}_{C'}(\tau,\xi)\rangle_{C'}T_A\bigg] e^{-i(k\cdot (\omega-X)-k'\cdot (X'-\omega'))}\nonumber \\*
   &\times\text{tr}\bigg[\phi(\xi')\langle\mathcal{W}_{\tilde{C}}(\xi',\tau')\rangle_{\tilde{C}} \phi(\tau')\langle\mathcal{W}_{\tilde{C}'}(\tau',\xi')\rangle_{\tilde{C}'}T_B \bigg]. \label{DBDB}
\end{align}

To proceed the calculation, we need to evaluate Wick's contraction of the Wilson lines. For simplicity, we are not going to provide a full expression for that contraction but rather evaluate Wick's contraction of their Taylor expansions keeping up to the first non-trivial terms which are
\begin{equation}
    \langle (1-\int_{C_1} \mathcal{A}(\omega)\cdot d\omega)(1-\int_{C_2} \mathcal{A}(\omega')\cdot d\omega') \rangle_\mathcal{A} =1+\int_{C_1}\int_{C_2} \langle \mathcal{A}_i(\omega)\mathcal{A}_j(\omega')\rangle_\mathcal{A}d\omega^id\omega'^j. \label{WL Wick}
\end{equation}
We will show that keeping the series expansion in this way is enough to provide a structure to reproduce the three-point interaction of the Wilson loop. Accordingly, let's consider the double integral on the right-hand side. Using (\ref{A prop su2}), one obtains
\begin{align}
    \int_{C_1}\int_{C_2} \langle \mathcal{A}_i(\omega)\mathcal{A}_j(\omega')\rangle_\mathcal{A}d\omega^id\omega'^j= \frac{i}{4}\frac{1}{\rvert\phi\rvert^2}\epsilon^{ACB}T_A \phi_C T_B \Big(n[C_1,C_2]\Big) \label{AACC}
\end{align}
where 
\begin{equation}
    n[C_1,C_2]=\int_{C_1}\int_{C_2} \epsilon_{ij} \delta^{2}(\omega-\omega') d\omega^id\omega'^j
\end{equation}
which counts the number of times the two curves intersect in an oriented way. To obtain (\ref{AACC}), we choose to work in the gauge choice such that the unit vector $\hat\phi$ is constant everywhere in the worldsheet interior $\Sigma$ but its value varies along the worldsheet boundary $\partial\Sigma$. If the two curves $C_1$ and $C_2$ lie themselves on the upper half plane with the boundaries on the x-axis whose endpoints are $b_1$ and $b_2$ respectively, according to \cite{Curry:2017cnu}, the average over $C_1$ and $C_2$ of $n[C_1,C_2]$ is
\begin{equation}
    \langle n[C_1,C_2] \rangle_{C_1,C_2}=\alpha\frac{b_1-b_2}{\vert b_1-b_2\vert}=\alpha \ \text{sign}(b_1-b_2)
\end{equation}
with proportional constant $\alpha$. Remind that the average over curve was defined in (\ref{curve avr}). The Wick's contraction of the curve-averaged Wilson loops reads
\begin{align}
    \langle \wick{  \c1{\mathcal{W}}_{C_1}(\xi,\tau)\rangle_{C_1}\langle \c1{\mathcal{W}}_{C_2}(\xi',\tau')\rangle_{C_2}}=1-i\frac{\alpha}{4}\frac{1}{\rvert\phi\rvert^2}\epsilon^{ABC}\phi_C T_A  T_B \ \text{sign}(b_1-b_2).\label{wick wilson line}
\end{align}
Remember that $\rvert\phi\rvert^2$ is constant throughout the worldsheet.

As a consequence, applying (\ref{wick wilson line}), one finds the expectation value of the integrand (\ref{1st integrand}) as
\begin{align}
    \bigg\langle\text{tr}\bigg[&\mathbb{B}_{k \ \mu}(\tau,\xi)\mathbb{C}^\mu_{-k }(\xi,\tau)\bigg]\text{tr}\bigg[\mathbb{B}_{k' \ \nu}(\tau',\xi')\mathbb{C}^{\nu}_{-k'}(\xi',\tau')\bigg]\bigg\rangle_\mathcal{A}\nonumber \\
    \ni& \frac{i}{4}\delta^2(\xi-\xi')e^{-i(k\cdot (X-\omega)+k'\cdot (X'-\omega'))}\epsilon^{ACB}\frac{\phi_C}{\rvert\phi\rvert^2}(\xi)\phi^D(\tau)\phi^E(\xi)\phi^F(\tau')\phi^G(\xi')\nonumber \\
    & \times \epsilon^{ab}\partial_a X^\mu(\xi)\partial'_b(\xi') X'^\nu \mathbb{P}_k(\dot{\omega})_\mu(\tau)\mathbb{P}_{k'}(\dot{\omega})_\nu(\tau')  \nonumber \\
    & \times \bigg\{ 2\text{tr}(T_D T_E T_A)\text{tr}(T_F T_G T_B)-i\frac{\alpha}{4}\frac{\epsilon^{IJK}\phi_K}{\rvert\phi\rvert^2}(\xi)\text{sign}(\omega-\omega')\Theta_1 \nonumber \\
    &\ -\frac{\alpha^2}{16}\frac{\epsilon^{IJK}\phi_K}{\rvert\phi\rvert^4}(\xi)\epsilon^{LMN}\phi_N(\xi)\Omega_1\bigg\}\label{BCBC2}
\end{align}
where 
\begin{align}
    \Theta_1= &\text{tr}(T_DT_ET_AT_I)\text{tr}(T_F T_G T_B T_J)+\text{tr}(T_DT_IT_ET_A)\text{tr}(T_F T_J T_G T_B) \nonumber \\
    &-\text{tr}(T_DT_IT_ET_A)\text{tr}(T_F T_G T_B T_J)-\text{tr}(T_DT_ET_AT_I)\text{tr}(T_F T_J T_G T_B)
\end{align}
and
\begin{align}
    \Omega_1= &\text{tr}(T_DT_IT_ET_AT_L)\text{tr}(T_F T_J T_G T_B T_M) \nonumber \\*
    &+\text{tr}(T_DT_IT_ET_AT_L)\text{tr}(T_F T_M T_G T_B T_J).
\end{align}
Similarly, we can find the expectation values of the integrand (\ref{2nd integrand}) and (\ref{3rd integrand}) as
\begin{align}
    \bigg\langle\text{tr}\bigg[&\mathbb{B}_{k \ \mu}(\tau,\xi)\mathbb{C}^\mu_{-k }(\xi,\tau)\bigg]\text{tr}\bigg[\mathbb{D}^{a\nu}_{k'}(\xi',\tau')\partial'_a\mathbb{B}_{-k' \ \nu}(\tau',\xi')\bigg]\bigg\rangle_\mathcal{A} \nonumber \\
    \ni& \frac{i}{4}\delta^2(\xi-\xi')e^{-i(k\cdot (X-\omega)-k'\cdot (X'-\omega'))}\epsilon^{ACB}\frac{\phi_C}{\vert\phi\vert^2}(\xi)\phi^D(\tau)\phi^E(\xi)\phi^F(\tau')\phi^G(\xi')   \nonumber \\
    & \times\epsilon^{ab}\partial_a X^\mu(\xi)\partial'_b(\xi') X'^\nu \mathbb{P}_k(\dot{\omega})_\mu(\tau)\mathbb{P}_{-k'}(\dot{\omega})_\nu(\tau')  \nonumber \\
    & \times \bigg\{ 2\text{tr}(T_D T_E T_A)\text{tr}(T_G T_F T_B) -i\frac{\alpha}{4}\frac{\epsilon^{IJK}\phi_K}{\vert\phi\vert^2}(\xi)\text{sign}(\omega-\omega')\Theta_2 \nonumber \\
    &-\frac{\alpha^2}{16}\frac{\epsilon^{IJK}\phi_K}{\vert\phi\vert^4}(\xi)\epsilon^{LMN}\phi_N(\xi)\Omega_2\bigg\}\label{BCDB2}
    \intertext{and}
        \bigg\langle \text{tr}\bigg[&\mathbb{D}^{a\mu}_{k}(\xi,\tau)\partial_a\mathbb{B}_{-k \ \mu}(\tau,\xi)\bigg]\text{tr}\bigg[\mathbb{D}^{b\nu}_{k'}(\xi',\tau')\partial'_b\mathbb{B}_{-k' \ \nu}(\tau',\xi')\bigg]\bigg\rangle_\mathcal{A} \nonumber \\
    \ni& \frac{i}{4}\delta^2(\xi-\xi')e^{-i(k\cdot (\omega-X)-k'\cdot (X'-\omega'))}\epsilon^{ACB}\frac{\phi_C}{\vert\phi\vert^2}(\xi)\phi^D(\tau)\phi^E(\xi)\phi^F(\tau')\phi^G(\xi')\nonumber \\
    & \times \epsilon^{ab}\partial_a X^\mu(\xi)\partial'_b(\xi') X'^\nu \mathbb{P}_{-k}(\dot{\omega})_\mu(\tau)\mathbb{P}_{-k'}(\dot{\omega})_\nu(\tau')  \nonumber \\
    & \times \bigg\{ 2\text{tr}(T_E T_D T_A)\text{tr}(T_G T_F T_B) -i\frac{\alpha}{4}\frac{\epsilon^{IJK}\phi_K}{\vert\phi\vert^2}(\xi)\text{sign}(\omega-\omega')\Theta_3 \nonumber \\
    &-\frac{\alpha^2}{16}\frac{\epsilon^{IJK}\phi_K}{\vert\phi\vert^4}(\xi)\epsilon^{LMN}\phi_N(\xi)\Omega_3\bigg\} \label{DBDB2}
\end{align}
where
\begin{align}
    \Theta_2= &\text{tr}(T_DT_IT_ET_A)\text{tr}(T_G T_F T_J T_B)+\text{tr}(T_DT_ET_AT_I)\text{tr}(T_G T_J T_F T_B) \nonumber \\
    &-\text{tr}(T_DT_ET_AT_I)\text{tr}(T_G T_F T_J T_B)-\text{tr}(T_DT_IT_ET_A)\text{tr}(T_G T_J T_F T_B), \\
    \Omega_2=& \text{tr}(T_DT_IT_ET_AT_L)\text{tr}(T_G T_J T_F T_M T_B)\nonumber \\
    &+\text{tr}(T_DT_IT_ET_AT_L)\text{tr}(T_G T_M T_F T_J T_B), \\
    \Theta_3= &\text{tr}(T_ET_IT_DT_A)\text{tr}(T_G T_J T_F T_B)+\text{tr}(T_ET_DT_IT_A)\text{tr}(T_G T_F T_J T_B) \nonumber \\
    &-\text{tr}(T_ET_DT_IT_A)\text{tr}(T_G T_J T_F T_B)-\text{tr}(T_ET_IT_DT_A)\text{tr}(T_G T_F T_J T_B), 
    \intertext{and}
        \Omega_3=& \text{tr}(T_ET_IT_DT_LT_A)\text{tr}(T_G T_J T_F T_M T_B)\nonumber \\
        &+\text{tr}(T_ET_IT_DT_LT_A)\text{tr}(T_G T_M T_F T_J T_B).
\end{align}

As we have everything set up, we shall proceed the calculation of the gauge field expectation of (\ref{3-pt ver SI}), $\Big\langle\frac{1}{2}(S_I^{\phi,\mathcal{A}})^2\Big\rangle_\mathcal{A}$, by substituting (\ref{BCBC2}), (\ref{BCDB2}), and (\ref{DBDB2}) into (\ref{3-pt ver SI}). This yields
\begin{align}
   &\Big\langle\frac{1}{2}(S_I^{\phi,\mathcal{A}})^2\Big\rangle_\mathcal{A}\ni 8iq^4\int_k \frac{1}{k^2}\int_{k'} \frac{1}{k'^2}\int_\Sigma d^2\xi\oint_{\partial\Sigma} d\tau\oint_{\partial\Sigma} d\tau' e^{-i[(k+k')\cdot X -k\cdot\omega-k'\cdot\omega']}\nonumber \\
   &\times\epsilon^{ACB}\frac{\phi_C}{\vert\phi\vert^2}(\xi)\phi^D(\tau)\phi^E(\xi)\phi^F(\tau')\phi^G(\xi) \Big(\epsilon^{ab}\partial_a X^\mu\partial_b X^\nu\Big)(\xi) \mathbb{P}_k(\dot{\omega})_\mu(\tau)\mathbb{P}_{k'}(\dot{\omega})_\nu(\tau') \nonumber \\
   & \times\bigg\{2\Big[\text{tr}(T_D T_E T_A)\text{tr}(T_F T_G T_B)+2\text{tr}(T_D T_E T_A)\text{tr}(T_G T_F T_B) \nonumber \\
   &+\text{tr}(T_E T_D T_A)\text{tr}(T_G T_F T_B)\Big]-i\frac{\alpha}{4}\frac{1}{\vert\phi\vert^2}\epsilon^{IJK}\phi_K(\xi)\text{sign}(\omega-\omega')\Big(\Theta_1+2\Theta_2+\Theta_3\Big) \nonumber \\ &-\frac{\alpha^2}{16}\frac{1}{\vert\phi\vert^4}\epsilon^{IJK}\phi_K(\xi)\epsilon^{LMN}\phi_N(\xi)\Big(\Omega_1+2\Omega_2+\Omega_3\Big)\bigg\}\label{3-pt ver SI2}
\end{align}
where $\int_k\equiv \int \frac{d^4k}{(2\pi)^4}$. To obtain the above equation, we utilized the fact that the integral $\int d^4k$ is invariant under a sign flip, i.e. $k \rightarrow -k$, thus one can rename the variables $k$ and $k'$ in (\ref{BCDB2}) and (\ref{DBDB2}) to have the same integrand regarding the matter field $X^\mu$ as in (\ref{BCBC2}). It is not difficult to see that the terms in the squared bracket in the third line sum up to zero as tr$(T_AT_BT_C)$ is all anti-symmetric in the case of $SU(2)$. The same situation also happens for the terms $\Theta_1+2\Theta_2+\Theta_3$ whose value sum up to zero. This can be seen using the expression for  a trace of a product of four generators \cite{10.21468/SciPostPhysLectNotes.21}, i.e.
\begin{align}
    \text{tr}(T_AT_BT_CT_D)=&\frac{1}{4N}(\delta_{AB}\delta_{CD}-\delta_{AC}\delta_{BD}+\delta_{AD}\delta_{BC})\nonumber \\
    &+\frac{1}{8}(d_{ABE}d_{CDE}-d_{ACE}d_{BDE}+d_{ADE}d_{BCE})\nonumber \\
    &+\frac{i}{8}(d_{ABE}f_{CDE}+d_{ACE}f_{BDE}+d_{ADE}f_{BCE}).
\end{align}
Remember that $d_{ABC}=0$ for $SU(2)$.

Therefore, what remains is to compute the integrand 
\begin{equation}
    \epsilon^{ABC}\phi_C(\xi)\phi^D(\tau)\phi^E(\xi)\phi^F(\tau')\phi^G(\xi)\epsilon^{IJK}\phi_K(\xi)\epsilon^{LMN}\phi_N(\xi)\Big(\Omega_1+2\Omega_2+\Omega_3\Big). \label{integrand last}
\end{equation}
To evaluate the quantities $\Omega$, we use a formula for trace of five $SU(2)$ generators
\begin{equation}
    \text{tr}(T_AT_BT_CT_DT_E)=\frac{i}{16}\delta_{AB}\epsilon_{CDE}+\frac{i}{16}\delta_{CD}\epsilon_{ABE}-\frac{i}{16}\delta_{BE}\epsilon_{ACD}+\frac{i}{16}\delta_{AE}\epsilon_{BCD}.  \label{trace 5}
\end{equation}
A more detailed derivation of the above relation can be found in Appendix \ref{trace gen}. As a consequence, the integrand (\ref{integrand last}) now takes the form
\begin{align}
    \Big(\frac{i}{16}\Big)^2\bigg[&(\phi^A(\tau)\mathcal{F}^{AB}\phi^B(\tau'))(\mathcal{F}^{CD}\mathcal{F}^{DE}\mathcal{F}^{EF}\mathcal{F}^{FC})\nonumber \\
    &+3(\phi^A(\tau)\mathcal{F}^{AB}\mathcal{F}^{BC}\mathcal{F}^{CD}\phi^D(\tau'))(\mathcal{F}^{IJ}\mathcal{F}^{JI}) \nonumber \\
    &-8(\phi^A(\tau)\mathcal{F}^{AB}\mathcal{F}^{BC}\mathcal{F}^{CD}\mathcal{F}^{DE}\mathcal{F}^{EF}\phi^F(\tau'))\Big] \label{integrand final}
\end{align}
where
\begin{equation}
    \mathcal{F}^{AB}=\epsilon^{ABC}\phi_C(\xi).
\end{equation}
We then further simplify (\ref{integrand final}) using the property of the Levi-Civita symbol which is 
\begin{equation}
    \epsilon^{ABC}\epsilon^{ADE}=\delta^{BD}\delta^{CE}-\delta^{BE}\delta^{CD}.
\end{equation}
Therefore, the integrand (\ref{integrand final}) becomes
\begin{equation}
    -4\Big(\frac{i}{16}\Big)^2 \phi^A(\tau)\phi^B(\tau')\epsilon^{ABC}\phi^C(\xi)\vert\phi\vert^4(\xi). \label{integrand final2}
\end{equation}
Substitute (\ref{integrand final2}) back to (\ref{3-pt ver SI2}), it yields
\begin{align}
   \Big\langle\frac{1}{2}(S_I^{\phi,\mathcal{A}})^2\Big\rangle_\mathcal{A}\ni& 2i\Big(\frac{q}{4}\Big)^4\alpha^2\int \frac{d^4k}{(2\pi)^4} \frac{1}{k^2}\int \frac{d^4k'}{(2\pi)^4} \frac{1}{k'^2}\int_\Sigma d^2\xi\oint_{\partial\Sigma} d\tau\oint_{\partial\Sigma} d\tau'\nonumber \\
   &\times\epsilon^{ABC}\phi^A(\tau)\phi^B(\tau')\frac{\phi^C}{\vert\phi\vert^2}(\xi) e^{-i[(k+k')\cdot X -k\cdot\omega-k'\cdot\omega']} \nonumber \\
   &\times \Big(\epsilon^{ab}\partial_a X^\mu\partial_b X^\nu\Big)(\xi) \mathbb{P}_k(\dot{\omega})_\mu(\tau)\mathbb{P}_{k'}(\dot{\omega})_\nu(\tau'). \label{3-pt ver SI3}
\end{align}

For the next step, we would like to apply the functional average for the field $X$ to the right-hand side of (\ref{3-pt ver SI3}). To do so, we will change the form of the bulk integral 
\begin{equation}
    \int_\Sigma d^2\xi \frac{\phi}{\vert\phi\vert^2}(\xi)\big(\epsilon^{ab}\partial_a X^\mu\partial_b X^\nu\big)(\xi) e^{-i(k+k')\cdot X}
\end{equation} 
into a new boundary vertex operator as we learned in the first section that the expectation inside the worldsheet gets greatly suppressed by an exponential of the Green's function at the co-incident points $G(\xi,\xi)$. This can be done by expanding the integral using the projection of $X$ (\ref{proj op}) along $(k+k')$. Therefore, there exists the term
\begin{equation}
   i \int_\Sigma d^2\xi \frac{\phi}{\vert\phi\vert^2}(\xi)\epsilon^{ab} \partial_a\mathbb{P}_{k+k'}(X)^\mu (k+k')^\nu \frac{1}{(k+k')^2}\partial_b e^{-i(k+k')
   \cdot X}
\end{equation}
in which we can turn it into an integral along the boundary using integration by
parts which is
\begin{equation}
    -i\oint_{\partial\Sigma} d\lambda \ \frac{\phi}{\vert\phi\vert^2}(\lambda) \frac{1}{(k+k')^2} \mathbb{P}_{k+k'}(\dot{\omega})^\mu(\lambda)(k+k')^\nu e^{-i(k+k')\cdot \omega(\lambda)} \label{projected bdry ver}
\end{equation}
where $\dot{\omega}^\mu(\lambda)=\frac{d\omega^\mu}{d\lambda}(\lambda)$. Substituting (\ref{projected bdry ver}) to (\ref{3-pt ver SI3}) and relabeling some dummy parameters, one obtains
\begin{align}
   \Big\langle\frac{1}{2}(S_I^{\phi,\mathcal{A}})^2\Big\rangle_{\mathcal{A},\Sigma}\ni& 2\Big(\frac{q}{4}\Big)^4\alpha^2\prod_{i=1}^3\int \frac{d^4k_i}{(2\pi)^4} (2\pi)^4\delta^4(k_1+k_2+k_3) \prod_{i=1}^3 \oint_{\partial\Sigma} d\tau_i \ \frac{1}{k_1^2k_2^2k_3^2} \nonumber \\
   &\times\epsilon^{ABC}\phi^A(\tau_1)\phi^B(\tau_2)\frac{\phi^C}{\vert\phi\vert^2}(\tau_3) e^{-i\sum_{i=1}^3 k_i\cdot \omega(\tau_i)} \nonumber \\
   &\times \mathbb{P}_{k_1}(\dot{\omega})_\mu(\tau_1)\mathbb{P}_{k_3}(\dot{\omega})^\mu(\tau_3)\mathbb{P}_{k_2}(\dot{\omega})_\nu(\tau_2)k_3^\nu. \label{3-pt ver SI4}
\end{align}

To relate the right-hand side of the above expression to the expectation of the Wilson loop with three-point self-interaction at the lowest order (\ref{WL 3 int}), we expect that the expectation of the product of $\phi$ would provide a notion of path-ordering of the trace of generators. As a consequence, what we need to do next is to determine the expectation of $\frac{1}{\vert\phi\vert^2} \phi^A\phi^B\phi^C$ concerning the remaining fields, i.e.
\begin{align}
    \Big\langle \phi^A(\tau_1)\phi^B(\tau_2)\frac{\phi^C}{\vert\phi\vert^2}(\tau_3)) \Big\rangle_{\phi,\psi^\dagger,\psi}. \label{3-pt exp}
\end{align}
According to (\ref{gen fn A wil WL}), a functional average of an object $\Xi$ with respect to the fields $\psi^\dagger,\psi,\phi$ is defined as
\begin{align}
    \langle \Xi \rangle_{\phi,\psi^\dagger,\psi}=\frac{1}{Z[\phi,\psi^\dagger,\psi]}\int D(\psi^\dagger,\psi,\phi)\prod_{i=1}^2 \frac{\delta(\mathcal{G}_i(\widetilde{\mathcal{J}}))}{\prod\limits_{\forall \xi} \text{det}(\widetilde{M})} \ \Xi \ e^{-S^{SU(2)}_\text{eff}(\phi)}e^{\int d\tau \psi^\dagger \dot{\psi}}
\end{align}
where 
\begin{equation}
   S_{\text{eff}}^{SU(2)}(\phi)= \frac{i}{2}\int_\Sigma d^2\xi \ \partial_i\phi\indices{_A}\partial_j\phi\indices{_B} \epsilon^{ij}\phi_C\epsilon\indices{^A^B^C}. \label{SU2 eff unit}
\end{equation}
and $\mathcal{G}_i(\widetilde{\mathcal{J}})$ is a boundary constraint defined in (\ref{boundary constraint}). The partition function $Z[\phi,\psi^\dagger,\psi]$ was placed to normalize the expectation which is defined as $Z[\phi,\psi^\dagger,\psi]=\langle 1 \rangle_{\phi,\psi^\dagger,\psi}$. Note that $\text{det}(\widetilde{M})=2\vert\phi\vert^2$ for $SU(2)$.

If we consider a worldsheet of which the topology is equivalent to that of a disk, we can map the worldsheet into an upper-half plane with a boundary on the x-axis. The expectation (\ref{3-pt exp}) becomes
\begin{align}
\frac{1}{Z[\phi,\psi^\dagger,\psi]} \int D(\psi^\dagger,\psi,\phi)&\  \delta(\text{tr}((2\partial_x\phi)\hat\phi))\delta(\text{tr}((2\partial_y\phi-2\phi\delta(y)-\widetilde{\mathcal{J}}_y)\hat\phi))\prod\limits_{\forall x}\frac{1}{2\vert\phi\vert^2} \nonumber \\
&\times \phi^A(x_1)\phi^B(x_2)\frac{\phi^C}{\vert\phi\vert^2}(x_3) e^{\int d\tau \psi^\dagger \dot{\psi}} \label{3-pt exp2}
\end{align}
where 
\begin{equation}
    \widetilde{\mathcal{J}}^A_y=\psi^\dagger(x) T^A \psi(x)\delta(y)
\end{equation}
and $x_i=x(\tau_i)$.
Remember that this path integral was taken along the worldsheet boundary on which the effective action $S_{\text{eff}}^{SU(2)}$ vanishes.

We found that if we imposed a boundary condition such that a derivative of the field $\phi$ in the normal direction equals zero, i.e. $\partial_y\phi=0$, the expectation (\ref{3-pt exp2}) gives path-ordering of the trace of the product of three generators. To illustrate this, let's consider the integral
\begin{equation}
    \int D\phi \ \delta(\text{tr}((2\partial_x\phi)\hat\phi))\delta(\text{tr}((2\phi\delta(y)+\widetilde{\mathcal{J}}_y)\hat\phi))
\end{equation}
where the boundary condition was already imposed. Taking a closer look at the constraints, the first constraint $\delta(\text{tr}((2\partial_x\phi)\hat\phi))$ implies that a modulus of the scalar field $\phi$ is constant throughout the boundary and the second one $\delta(\text{tr}((2\phi\delta(y)+\widetilde{\mathcal{J}}_y)\hat\phi))$ tells us that $\hat\phi(x)$ is required to equal to $\frac{1}{2\vert\phi\vert}\psi^\dagger(x) T \psi(x)$. Note that we are not interested in the solution where $\hat\phi=0$.

Accordingly, we proceed with the calculation by changing the variables of the measure $D\phi$ and the Dirac deltas to be
\begin{align}
    D\phi &=\prod\limits_{\forall x} \vert\phi\vert^2 d\vert\phi\vert d\hat\phi,\\
    \delta^{(3)}(\text{tr}((2\partial_x\phi)\hat\phi)) &= \int_0^\infty dc\ \frac{\delta(\vert\phi\vert-c)}{J_1}  ,\\
    \delta^{(3)}(\text{tr}((2\phi\delta(y)+\widetilde{\mathcal{J}}_y)\hat\phi)) &= \frac{\delta(\hat\phi+\frac{1}{2\vert\phi\vert}\psi^\dagger T \psi)}{J_2} 
\end{align}
with the Jacobian factors $J_1$ and $J_2$. It is not difficult to check that both $J_1$ and $J_2$ do not depend on either $\vert\phi\vert$ or $\hat\phi$. Exploiting this fact, it is not necessary to find the exact forms of the Jacobian factors as they will eventually cancel out with ones in $Z[\phi,\psi^\dagger,\psi]$. Consequently, (\ref{3-pt exp2}) takes the form
\begin{align}
\frac{1}{\widetilde{Z}[\phi,\psi^\dagger,\psi]} \int D(\psi^\dagger,\psi)&\prod\limits_{\forall x} \bigg( \vert\phi\vert^2 d\vert\phi\vert d\hat\phi\  \int_0^\infty dc\ \delta(\vert\phi\vert-c)\delta(\hat\phi+\frac{1}{2\vert\phi\vert}\psi^\dagger T \psi)  \frac{1}{2\vert\phi\vert^2} \bigg) \nonumber \\
&\times \phi^A(x_1)\phi^B(x_2)\frac{\phi^C}{\vert\phi\vert^2}(x_3) e^{\int d\tau \psi^\dagger \dot{\psi}} \label{3-pt exp3}
\end{align}
where the Jacobian factors $J_1$ and $J_2$ have been omitted as discussed and $\widetilde{Z}[\phi,\psi^\dagger,\psi]$ takes the form
\begin{equation}
    \widetilde{Z}[\phi,\psi^\dagger,\psi]=\int D(\psi^\dagger,\psi)\prod\limits_{\forall x} \bigg( \frac{1}{2} \int_0^\infty dc \bigg)e^{\int d\tau \psi^\dagger \dot{\psi}}. \label{Z denominator}
\end{equation}
By integrating out the unit scalar field $\hat\phi$, (\ref{3-pt exp3}) becomes
\begin{align}
\frac{1}{\widetilde{Z}[\phi,\psi^\dagger,\psi]} \int &D(\psi^\dagger,\psi)\Bigg[\prod\limits_{\forall x\neq x_3} \bigg( \frac{1}{2} d\vert\phi\vert  \int_0^\infty dc\ \delta(\vert\phi\vert-c) \bigg)\Bigg] \nonumber \\
&\times\bigg( \frac{1}{2} d\vert\phi\vert \frac{1}{\vert\phi\vert^2} \int_0^\infty dc\ \delta(\vert\phi\vert-c) \bigg)\Bigg\vert_{x_3} e^{\int d\tau \psi^\dagger \dot{\psi}} \nonumber \\
&\times\frac{1}{8}\psi^\dagger(x_1) T^A \psi(x_1)\psi^\dagger(x_2) T^B \psi(x_2) \psi^\dagger(x_3) T^C \psi(x_3) . \label{3-pt exp4}
\end{align}
Notice that the product in the square bracket is evaluated at every point on the boundary except for $x=x_3$ while the integrand in the second line is at $x=x_3$. It is obvious that the integral over $\vert\phi\vert$ in the first line will get canceled out by the same integrals in $\widetilde{Z}[\phi,\psi^\dagger,\psi]$ for every point $x\neq x_3$. However, at the point $x_3$, the integral over $\vert\phi\vert$ reads
\begin{align}
   \frac{1}{2} \int_0^\infty d\vert\phi\vert \frac{1}{\vert\phi\vert^2} \int_0^\infty dc\ \delta(\vert\phi\vert-c)=\frac{1}{2}\int_0^\infty dc\ \frac{1}{c^2}.
\end{align}
This is exactly the integral in the parenthesis of (\ref{Z denominator}) when relabelling $c \rightarrow \frac{1}{c}$ so they also cancel out. In consequence, this turns the expectation (\ref{3-pt exp}) to be
\begin{equation}
   \frac{1}{8} \int D(\psi^\dagger,\psi) \ \psi^\dagger(x_1) T^A \psi(x_1)\psi^\dagger(x_2) T^B \psi(x_2) \psi^\dagger(x_3) T^C \psi(x_3)  e^{\int d\tau \psi^\dagger \dot{\psi}}
\end{equation}
which can be interpreted as 
\begin{equation}
    \frac{1}{8} \mathcal{P}\text{tr}\Big( T^A T^B T^C \Big)
\end{equation}
according to \cite{Samuel:1978iy,Broda:1995wv}.

All things considered, we find that the expectation value of $(S_I^{\phi,\mathcal{A}})^2$ in the proposed string model contains a term
\begin{align}
   \Big\langle\frac{1}{2}(S_I^{\phi,\mathcal{A}})^2\Big\rangle_{\mathcal{A},\Sigma}\ni& \Big(\frac{q}{4}\Big)^4\frac{\alpha^2}{4}\prod_{i=1}^3\int \frac{d^4k_i}{(2\pi)^4} \prod_{i=1}^3 \oint_{\partial\Sigma} d\tau_i \  (2\pi)^4\delta^4(k_1+k_2+k_3) \nonumber \\*
   &\times \epsilon^{ABC} \mathcal{P}\text{tr}\Big( T^A T^B T^C \Big)\frac{1}{k_1^2k_2^2k_3^2} e^{-i\sum_{i=1}^3 k_i\cdot \omega(\tau_i)}
   \nonumber \\*
   &\times\mathbb{P}_{k_1}(\dot{\omega})_\mu(\tau_1)\mathbb{P}_{k_3}(\dot{\omega})^\mu(\tau_3)\mathbb{P}_{k_2}(\dot{\omega})_\nu(\tau_2)k_3^\nu  \label{3-pt ver SI5}
\end{align}
which is the form we need to relate to the expectation value in
the gauge theory of the Wilson loop at the $\mathcal{O}(q^4)$ (\ref{WL 3 int}). Nevertheless, we may leave the constant $\alpha$ undetermined as it is probable that similar expressions can also be obtained when including higher order of corrections in Wick's contraction of Wilson loops. Remember that to obtain the expression (\ref{3-pt ver SI5}), we only keep Wick's contraction of the Wilson loops up to the first non-trivial term. 

It is undeniable that the computation presented in section five relied heavily on the properties of $SU(2)$ group generators. To generalize to $SU(N)$, one needs to handle more complicated relations among $SU(N)$ group generators, which involve both structure constants $f^{ABC}$ and totally symmetric third-rank tensors $d^{ABC}$. We speculate that the three-point vertex could be reproduced from the expectation of the same term in equation (\ref{3-pt ver SI}).

However, the model is still incomplete without a reproduction of four-point self-interactions. We speculate that the four-point vertex term could be obtained from the expectation value of $(S_I^{\phi,\mathcal{A}})^3$ as the four-gluon vertex is first observed at the order $q^6$. Needless to say, the computation would be significantly more difficult and will need to be left for future investigation.

\section{Concluding Remarks}\label{sec6}
To conclude, we proposed a possible non-Abelian generalization of the \cite{Edwards:2014cga,Edwards:2014xfa} via the introduction of Lie algebra-valued worldsheet fields into the string model. The dynamics of the new degrees of freedom are described by the topological BF action. In our model, the string theory has unusual contact interactions defined in (\ref{new model}) which enjoy local gauge symmetries. By inserting a Wilson loop along the worldsheet boundary, we showed that a boundary path-ordering of Lie generators can be reformulated through the expectation in equation (\ref{gauge inv obs}).

The partition function of the modified tensionless string model with contact interactions was presented in (\ref{Psi}). When neglecting the effects of the bulk terms, the partition function is equal to the expectation of a non-Abelian Wilson loop in the usual gauge theory when self-interactions have been turned off.

We then verified that our string model correctly reproduces the three-point self-interaction of $SU(2)$ Yang-Mills theory. We performed this calculation by perturbatively expanding the exponential of the contact interaction up to the second order. We found that at that order, the expectation of $(S_I^{\phi,\mathcal{A}})^2$ contains a precise expression for the expectation value of the $SU(2)$ Wilson loop with three-point interaction at the lowest order. To obtain this result, we used a worldsheet gauge propagator calculated in the section four.

Despite encouraging results, we cannot rush to the conclusion that our string model provides a true description of non-Abelian Yang-Mills theory. There are still a few issues to be further considered. Firstly, we have not yet investigated whether the model can reproduce a four-point interaction, and if it does, we expect that such a vertex can be found from the expectation of $(S_I^{\phi,\mathcal{A}})^3$. More importantly, it is still unexplored whether the model reconciles with general Lie algebras. Secondly, in this work, we ignore potential divergences that may occur when performing functional integration over the matter field $X$ when any two string vertex operators are placed very close to each other inside the worldsheet. In the Abelian case, this issue was addressed by including supersymmetry on the worldsheet \cite{Edwards:2014cga,Edwards:2014xfa}. Developing a supersymmetric generalization of the tensionless string model to describe non-Abelian super Yang-Mills is a subject for future investigation. Finally, it is also possible that our model will generate additional interacting vertices not found from the Wilson loop in the non-Abelian theory, which could imply potential deviation between the two models.

\section*{Acknowledgement}
We are pleased to acknowledge Paul Mansfield and Chris Curry for sharing the results and a manuscript of their ongoing work \cite{PM} with us in which parts of this work were greatly inspired. Moreover, we are thankful to Paul Mansfield for useful discussions.

\newpage
\begin{appendices}
\section{Non-Abelian Wilson loop of the Yang-Mills Theory} \label{non AB WL}
The Wilson loop can be defined as the trace of the path-ordered exponential of a line integral of the gauge field $A$ along a closed loop $C$,
\begin{equation}
    W[C]=\text{tr}\big(\mathcal{P}\big(e^{-q\oint_C A\cdot d\xi}\big)\big) \label{Wilson loop}
\end{equation}
where $\mathcal{P}$ is a path-ordering operator which orders a product of operators along the path $C$ as 
\begin{equation}
    \mathcal{P}(O_1(\xi_1)O_1(\xi_2) \ldots O_N(\xi_N)\equiv O_{P_1}(\xi_{P_1})O_{P_2}(\xi_{P_2}) \ldots O_{P_N}(\xi_{P_N})
\end{equation}
where on the right hand side the operators are ordered according to their positions along the path, i.e. $\xi_{P_1}\geq \xi_{P_2} \geq \ldots \geq \xi_{P_N}$. The trace in (\ref{Wilson loop})  is computed over colour indices.

By Taylor expanding the exponential  (\ref{Wilson loop}), the expectation value of the Wilson loop is 
\begin{equation}
    \langle W[C]\rangle=\text{tr}\bigg( \mathcal{P}\sum_{n=0}^\infty\frac{(-q)^n}{n!} \Big\langle \prod_{i=0}^n \oint_C d\xi_i^{\mu_i} A_{\mu_i} \Big\rangle \bigg). \label{exp wilson}
\end{equation}
For a small coupling constant $q$, we can evaluate (\ref{exp wilson}) perturbatively which requires the calculation of $\langle A^n\rangle$. The first non-trivial contribution to $\langle W \rangle$ is
\begin{equation}
    \frac{q^2}{2}\text{tr}\bigg( \mathcal{P} \oint_C \oint_C d\xi_1^\mu d\xi_2^\nu \big\langle A_\mu(\xi_1)A_\nu(\xi_2) \big\rangle \bigg).
\end{equation}
The above term can be written as
\begin{equation}
    \frac{q^2}{2}\text{tr}\bigg( \mathcal{P} \oint_C \oint_C d\xi_1^\mu d\xi_2^\nu \frac{d^4k}{(2\pi)^4}\eta_{AB}\bigg(\eta^{\mu\nu}-(1-\zeta)\frac{k^\mu k^\nu}{k^2}\bigg)\frac{e^{ik\cdot(\xi_1-\xi_2)}}{k^2} T^A T^B\bigg). \label{1 MW WL}
\end{equation}

If we ignore the self-interaction of the Yang-Mills theory, then the expectation value of the Wilson loop is evaluated as 
\begin{equation}
    \text{tr}\mathcal{P} \text{exp}\bigg( \frac{q^2}{2} \oint_C \oint_C d\xi_1^\mu d\xi_2^\nu \frac{d^4k}{(2\pi)^4}\eta_{AB}\bigg(\eta^{\mu\nu}-(1-\zeta)\frac{k^\mu k^\nu}{k^2}\bigg)\frac{e^{ik\cdot(\xi_1-\xi_2)}}{k^2} T^A T^B\bigg) \label{non-self-int WL}
\end{equation}
which only differs from the Abelian case by the path-ordering of the Lie generators.

The main difference between the Abelian and non-Abelian theories is the existence of self-interactions. There are three and four gauge boson vertices which first appear in the expectation of the Wilson loop at  $q^4$ and $q^6$ respectively. For the three-point vertex, we can calculate its contribution to the expectation of the Wilson loop at the lowest order by considering
\begin{equation}
    \text{tr}\mathcal{P} \bigg( -\frac{q^3}{3!} \langle (\oint d\xi^\mu A_\mu)^3\rangle \bigg)=  -\frac{q^3}{3!}\frac{1}{Z[0]} \text{tr}\mathcal{P}  \prod_{i=1}^3 \bigg( \oint d\xi_i^\mu T^{A_i}\frac{\delta}{\delta J^{\mu A_i}(\xi_i)}\bigg) Z[J,\eta,\bar{\eta}]\bigg\vert_{J,\eta,\bar{\eta}=0}
\end{equation}
where $Z[J,\eta,\bar{\eta}]$ is a generating functional defined as
\begin{equation}
    Z[J,\eta,\bar{\eta}]=\int DAD\bar{c}Dc \exp \bigg[-\int d^4x \bigg(\mathcal{L}_\text{g.f.}-J^R_\mu A^
    \mu_R -\bar{\eta}_R c^R-\eta_R \bar{c}^R\bigg) \bigg]. \label{generating functional}
\end{equation}
The gauge-fixed Yang-Mills Lagrangian $\mathcal{L}_\text{g.f.}$ is written as 
\begin{equation}
    \mathcal{L}_{\text{g.f.}}(x)=-\frac{1}{4}F^R_{\mu\nu}F_R^{\mu\nu}-\frac{(\partial^\mu A_\mu)^2}{2\zeta}-\bar{c}_A(\partial^\mu \mathcal{D}^{AB}_\mu)c_B. \label{gf YM Lagrangian}
\end{equation}
Keeping only the terms with only $f^{ABC}$, one obtains
\begin{gather}
    \text{tr}\mathcal{P} \bigg[i\frac{q^4}{2}f^{ABC}T_A T_B T_C \prod_{i=1}^3\int \frac{d^4k_i}{(2\pi)^4} (2\pi)^4\delta^4(k_1+k_2+k_3) \oint\oint\oint \frac{1}{k_1^2 k_2^2 k_3^2} \nonumber \\
    \times \bigg( d\xi_1^\mu-\frac{k_1^\mu k_1\cdot d\xi_1}{k_1^2} \bigg)\bigg( d\xi_{2\mu}-\frac{k_{2\mu} k_2\cdot d\xi_2}{k_2^2} \bigg) ik_{1\nu}\bigg( d\xi_3^\nu-\frac{k_3^\nu k_3\cdot d\xi_3}{k_3^2} \bigg) e^{-i\sum_{i=1}^3 k_i\cdot \xi_i}. \label{WL 3 int}
\end{gather}
The expression (\ref{WL 3 int}) was computed in the Landau gauge ($\zeta=0$).

\section{Trace of Five \texorpdfstring{$SU(2)$}{SU(2)} Generators}\label{trace gen}
Consider the generators of $SU(N)$, denoted by $T_A$, with the normalization 
\begin{equation}
    \text{tr}(T_AT_B)=\frac{1}{2}\eta_{AB}.
\end{equation}
The product of the two generators is written as
\begin{equation}
    T_AT_B=\frac{1}{2}\bigg(\frac{1}{N}\eta_{AB}+\Big(d_{ABC}+if_{ABC}\Big)T^C\bigg) \label{TT}
\end{equation}
where $d_{ABC}$ is a totally symmetric tensor expressed by
\begin{equation}
    d_{ABC}=2\text{tr}(\{T_A,T_B\}T_C)
\end{equation}
and $f_{ABC}$ is a structure constant defined as
\begin{equation}
    [T_A,T_B]=if_{ABC}T^C.
\end{equation}
From (\ref{TT}), one finds
\begin{align}
    T_AT_BT_C=&\frac{1}{2N}\eta_{AB}T_C+\frac{1}{4N}\Big(d_{ABC}+if_{ABC}\Big) \nonumber \\
    &+\frac{1}{4}\Big(d\indices{_{AB}^D}+if\indices{_{AB}^D}\Big)\Big(d_{DCE}+if_{DCE}\Big)T^E.
\end{align}
In the case of $SU(2)$, if we identify $T^A=\frac{1}{2}\sigma^A$ where $\sigma^A$ are the Pauli matrices, the metric $\eta_{AB}$ is simply $\delta_{AB}$, thus there is no distinction between upper and lower indices. Moreover, in this scenario, $f_{ABC}=\epsilon_{ABC}$ and $d_{ABC}=0$. As a consequence, one obtains
\begin{equation}
    \text{tr}(T_AT_BT_CT_DT_E)=\frac{i}{16}\delta_{AB}\epsilon_{CDE}+\frac{i}{16}\delta_{CD}\epsilon_{ABE}-\frac{i}{16}\delta_{BE}\epsilon_{ACD}+\frac{i}{16}\delta_{AE}\epsilon_{BCD}. \label{trace 5 app}
\end{equation}
Remember that the generator $T^A$ is traceless. Readers may consult \cite{10.21468/SciPostPhysLectNotes.21,DEAZCARRAGA1998657} for more detailed discussion on relations among the $SU(N)$ generators.

\end{appendices}

\bibliography{sn-bibliography}


\begin{thebibliography}{32}
\ifx \bisbn   \undefined \def \bisbn  #1{ISBN #1}\fi
\ifx \binits  \undefined \def \binits#1{#1}\fi
\ifx \bauthor  \undefined \def \bauthor#1{#1}\fi
\ifx \batitle  \undefined \def \batitle#1{#1}\fi
\ifx \bjtitle  \undefined \def \bjtitle#1{#1}\fi
\ifx \bvolume  \undefined \def \bvolume#1{\textbf{#1}}\fi
\ifx \byear  \undefined \def \byear#1{#1}\fi
\ifx \bissue  \undefined \def \bissue#1{#1}\fi
\ifx \bfpage  \undefined \def \bfpage#1{#1}\fi
\ifx \blpage  \undefined \def \blpage #1{#1}\fi
\ifx \burl  \undefined \def \burl#1{\textsf{#1}}\fi
\ifx \doiurl  \undefined \def \doiurl#1{\url{https://doi.org/#1}}\fi
\ifx \betal  \undefined \def \betal{\textit{et al.}}\fi
\ifx \binstitute  \undefined \def \binstitute#1{#1}\fi
\ifx \binstitutionaled  \undefined \def \binstitutionaled#1{#1}\fi
\ifx \bctitle  \undefined \def \bctitle#1{#1}\fi
\ifx \beditor  \undefined \def \beditor#1{#1}\fi
\ifx \bpublisher  \undefined \def \bpublisher#1{#1}\fi
\ifx \bbtitle  \undefined \def \bbtitle#1{#1}\fi
\ifx \bedition  \undefined \def \bedition#1{#1}\fi
\ifx \bseriesno  \undefined \def \bseriesno#1{#1}\fi
\ifx \blocation  \undefined \def \blocation#1{#1}\fi
\ifx \bsertitle  \undefined \def \bsertitle#1{#1}\fi
\ifx \bsnm \undefined \def \bsnm#1{#1}\fi
\ifx \bsuffix \undefined \def \bsuffix#1{#1}\fi
\ifx \bparticle \undefined \def \bparticle#1{#1}\fi
\ifx \barticle \undefined \def \barticle#1{#1}\fi
\bibcommenthead
\ifx \bconfdate \undefined \def \bconfdate #1{#1}\fi
\ifx \botherref \undefined \def \botherref #1{#1}\fi
\ifx \url \undefined \def \url#1{\textsf{#1}}\fi
\ifx \bchapter \undefined \def \bchapter#1{#1}\fi
\ifx \bbook \undefined \def \bbook#1{#1}\fi
\ifx \bcomment \undefined \def \bcomment#1{#1}\fi
\ifx \oauthor \undefined \def \oauthor#1{#1}\fi
\ifx \citeauthoryear \undefined \def \citeauthoryear#1{#1}\fi
\ifx \endbibitem  \undefined \def \endbibitem {}\fi
\ifx \bconflocation  \undefined \def \bconflocation#1{#1}\fi
\ifx \arxivurl  \undefined \def \arxivurl#1{\textsf{#1}}\fi
\csname PreBibitemsHook\endcsname

\bibitem{Neveu:1971mu}
\begin{barticle}
\bauthor{\bsnm{Neveu}, \binits{A.}},
\bauthor{\bsnm{Scherk}, \binits{J.}}:
\batitle{{Connection between Yang-Mills fields and dual models}}.
\bjtitle{Nucl. Phys. B}
\bvolume{36},
\bfpage{155}--\blpage{161}
(\byear{1972}).
\doiurl{10.1016/0550-3213(72)90301-X}
\end{barticle}
\endbibitem

\bibitem{Kawai:1985xq}
\begin{barticle}
\bauthor{\bsnm{Kawai}, \binits{H.}},
\bauthor{\bsnm{Lewellen}, \binits{D.C.}},
\bauthor{\bsnm{Tye}, \binits{S.H.H.}}:
\batitle{{A Relation Between Tree Amplitudes of Closed and Open Strings}}.
\bjtitle{Nucl. Phys. B}
\bvolume{269},
\bfpage{1}--\blpage{23}
(\byear{1986}).
\doiurl{10.1016/0550-3213(86)90362-7}
\end{barticle}
\endbibitem

\bibitem{Stieberger:2022lss}
\begin{botherref}
\oauthor{\bsnm{Stieberger}, \binits{S.}}:
{A Relation between One-Loop Amplitudes of Closed and Open Strings (One-Loop
  KLT Relation)}
(2022)
{\href{https://arxiv.org/abs/2212.06816}{{arXiv:2212.06816}}}
{[hep-th]}
\end{botherref}
\endbibitem

\bibitem{PhysRevD.78.085011}
\begin{barticle}
\bauthor{\bsnm{Bern}, \binits{Z.}},
\bauthor{\bsnm{Carrasco}, \binits{J.J.M.}},
\bauthor{\bsnm{Johansson}, \binits{H.}}:
\batitle{New relations for gauge-theory amplitudes}.
\bjtitle{Phys. Rev. D}
\bvolume{78},
\bfpage{085011}
(\byear{2008}).
\doiurl{10.1103/PhysRevD.78.085011}
\end{barticle}
\endbibitem

\bibitem{Stieberger:2009hq}
\begin{botherref}
\oauthor{\bsnm{Stieberger}, \binits{S.}}:
{Open \& Closed vs. Pure Open String Disk Amplitudes}
(2009)
{\href{https://arxiv.org/abs/0907.2211}{{arXiv:0907.2211}}}
{[hep-th]}
\end{botherref}
\endbibitem

\bibitem{Bjerrum-Bohr:2009ulz}
\begin{barticle}
\bauthor{\bsnm{Bjerrum-Bohr}, \binits{N.E.J.}},
\bauthor{\bsnm{Damgaard}, \binits{P.H.}},
\bauthor{\bsnm{Vanhove}, \binits{P.}}:
\batitle{{Minimal Basis for Gauge Theory Amplitudes}}.
\bjtitle{Phys. Rev. Lett.}
\bvolume{103},
\bfpage{161602}
(\byear{2009})
{\href{https://arxiv.org/abs/0907.1425}{{arXiv:0907.1425}}}
{[hep-th]}.
\doiurl{10.1103/PhysRevLett.103.161602}
\end{barticle}
\endbibitem

\bibitem{Cachazo:2013gna}
\begin{barticle}
\bauthor{\bsnm{Cachazo}, \binits{F.}},
\bauthor{\bsnm{He}, \binits{S.}},
\bauthor{\bsnm{Yuan}, \binits{E.Y.}}:
\batitle{{Scattering equations and Kawai-Lewellen-Tye orthogonality}}.
\bjtitle{Phys. Rev. D}
\bvolume{90}(\bissue{6}),
\bfpage{065001}
(\byear{2014})
{\href{https://arxiv.org/abs/1306.6575}{{arXiv:1306.6575}}}
{[hep-th]}.
\doiurl{10.1103/PhysRevD.90.065001}
\end{barticle}
\endbibitem

\bibitem{Stieberger:2016lng}
\begin{barticle}
\bauthor{\bsnm{Stieberger}, \binits{S.}},
\bauthor{\bsnm{Taylor}, \binits{T.R.}}:
\batitle{{New relations for Einstein\textendash{}Yang\textendash{}Mills
  amplitudes}}.
\bjtitle{Nucl. Phys. B}
\bvolume{913},
\bfpage{151}--\blpage{162}
(\byear{2016})
{\href{https://arxiv.org/abs/1606.09616}{{arXiv:1606.09616}}}
{[hep-th]}.
\doiurl{10.1016/j.nuclphysb.2016.09.014}
\end{barticle}
\endbibitem

\bibitem{Srisangyingcharoen:2020lhx}
\begin{barticle}
\bauthor{\bsnm{Srisangyingcharoen}, \binits{P.}},
\bauthor{\bsnm{Mansfield}, \binits{P.}}:
\batitle{{Plahte Diagrams for String Scattering Amplitudes}}.
\bjtitle{JHEP}
\bvolume{04},
\bfpage{017}
(\byear{2021})
{\href{https://arxiv.org/abs/2005.01712}{{arXiv:2005.01712}}}
{[hep-th]}.
\doiurl{10.1007/JHEP04(2021)017}
\end{barticle}
\endbibitem

\bibitem{tHooft:1973alw}
\begin{barticle}
\bauthor{\bparticle{'t} \bsnm{Hooft}, \binits{G.}}:
\batitle{{A Planar Diagram Theory for Strong Interactions}}.
\bjtitle{Nucl. Phys. B}
\bvolume{72},
\bfpage{461}
(\byear{1974}).
\doiurl{10.1016/0550-3213(74)90154-0}
\end{barticle}
\endbibitem

\bibitem{Edwards:2014cga}
\begin{barticle}
\bauthor{\bsnm{Edwards}, \binits{J.P.}},
\bauthor{\bsnm{Mansfield}, \binits{P.}}:
\batitle{{QED as the tensionless limit of the spinning string with contact
  interaction}}.
\bjtitle{Phys. Lett. B}
\bvolume{746},
\bfpage{335}--\blpage{340}
(\byear{2015})
{\href{https://arxiv.org/abs/1409.4948}{{arXiv:1409.4948}}}
{[hep-th]}.
\doiurl{10.1016/j.physletb.2015.05.024}
\end{barticle}
\endbibitem

\bibitem{Edwards:2014xfa}
\begin{barticle}
\bauthor{\bsnm{Edwards}, \binits{J.P.}},
\bauthor{\bsnm{Mansfield}, \binits{P.}}:
\batitle{{Delta-function Interactions for the Bosonic and Spinning Strings and
  the Generation of Abelian Gauge Theory}}.
\bjtitle{JHEP}
\bvolume{01},
\bfpage{127}
(\byear{2015})
{\href{https://arxiv.org/abs/1410.3288}{{arXiv:1410.3288}}}
{[hep-th]}.
\doiurl{10.1007/JHEP01(2015)127}
\end{barticle}
\endbibitem

\bibitem{Edwards:2015hka}
\begin{barticle}
\bauthor{\bsnm{Edwards}, \binits{J.P.}}:
\batitle{{Contact interactions between particle worldlines}}.
\bjtitle{JHEP}
\bvolume{01},
\bfpage{033}
(\byear{2016})
{\href{https://arxiv.org/abs/1506.08130}{{arXiv:1506.08130}}}
{[hep-th]}.
\doiurl{10.1007/JHEP01(2016)033}
\end{barticle}
\endbibitem

\bibitem{Mansfield:2011eq}
\begin{barticle}
\bauthor{\bsnm{Mansfield}, \binits{P.}}:
\batitle{{Faraday's Lines of Force as Strings: from Gauss' Law to the Arrow of
  Time}}.
\bjtitle{JHEP}
\bvolume{10},
\bfpage{149}
(\byear{2012})
{\href{https://arxiv.org/abs/1108.5094}{{arXiv:1108.5094}}}
{[hep-th]}.
\doiurl{10.1007/JHEP10(2012)149}
\end{barticle}
\endbibitem

\bibitem{Kalb:1974yc}
\begin{barticle}
\bauthor{\bsnm{Kalb}, \binits{M.}},
\bauthor{\bsnm{Ramond}, \binits{P.}}:
\batitle{{Classical direct interstring action}}.
\bjtitle{Phys. Rev. D}
\bvolume{9},
\bfpage{2273}--\blpage{2284}
(\byear{1974}).
\doiurl{10.1103/PhysRevD.9.2273}
\end{barticle}
\endbibitem

\bibitem{Curry:2017cnu}
\begin{barticle}
\bauthor{\bsnm{Curry}, \binits{C.}},
\bauthor{\bsnm{Mansfield}, \binits{P.}}:
\batitle{{Intersection of world-lines on curved surfaces and path-ordering of
  the Wilson loop}}.
\bjtitle{JHEP}
\bvolume{06},
\bfpage{081}
(\byear{2018})
{\href{https://arxiv.org/abs/1712.04760}{{arXiv:1712.04760}}}
{[hep-th]}.
\doiurl{10.1007/JHEP06(2018)081}
\end{barticle}
\endbibitem

\bibitem{DeWitt:1965jb}
\begin{barticle}
\bauthor{\bsnm{DeWitt}, \binits{B.S.}}:
\batitle{{Dynamical theory of groups and fields}}.
\bjtitle{Conf. Proc. C}
\bvolume{630701},
\bfpage{585}--\blpage{820}
(\byear{1964})
\end{barticle}
\endbibitem

\bibitem{McAvity:1990we}
\begin{barticle}
\bauthor{\bsnm{McAvity}, \binits{D.M.}},
\bauthor{\bsnm{Osborn}, \binits{H.}}:
\batitle{{A DeWitt expansion of the heat kernel for manifolds with a
  boundary}}.
\bjtitle{Class. Quant. Grav.}
\bvolume{8},
\bfpage{603}--\blpage{638}
(\byear{1991}).
\doiurl{10.1088/0264-9381/8/4/008}
\end{barticle}
\endbibitem

\bibitem{McAvity:1991xf}
\begin{barticle}
\bauthor{\bsnm{McAvity}, \binits{D.M.}},
\bauthor{\bsnm{Osborn}, \binits{H.}}:
\batitle{{Asymptotic expansion of the heat kernel for generalized boundary
  conditions}}.
\bjtitle{Class. Quant. Grav.}
\bvolume{8},
\bfpage{1445}--\blpage{1454}
(\byear{1991}).
\doiurl{10.1088/0264-9381/8/8/010}
\end{barticle}
\endbibitem

\bibitem{horowitz1989}
\begin{barticle}
\bauthor{\bsnm{Horowitz}, \binits{G.T.}}:
\batitle{Exactly soluble diffeomorphism invariant theories}.
\bjtitle{Comm. Math. Phys.}
\bvolume{125}(\bissue{3}),
\bfpage{417}--\blpage{437}
(\byear{1989})
\end{barticle}
\endbibitem

\bibitem{Blau:1989bq}
\begin{barticle}
\bauthor{\bsnm{Blau}, \binits{M.}},
\bauthor{\bsnm{Thompson}, \binits{G.}}:
\batitle{{Topological Gauge Theories of Antisymmetric Tensor Fields}}.
\bjtitle{Annals Phys.}
\bvolume{205},
\bfpage{130}--\blpage{172}
(\byear{1991}).
\doiurl{10.1016/0003-4916(91)90240-9}
\end{barticle}
\endbibitem

\bibitem{witten1991}
\begin{barticle}
\bauthor{\bsnm{Witten}, \binits{E.}}:
\batitle{On quantum gauge theories in two dimensions}.
\bjtitle{Comm. Math. Phys.}
\bvolume{141}(\bissue{1}),
\bfpage{153}--\blpage{209}
(\byear{1991})
\end{barticle}
\endbibitem

\bibitem{Witten:1992xu}
\begin{barticle}
\bauthor{\bsnm{Witten}, \binits{E.}}:
\batitle{{Two-dimensional gauge theories revisited}}.
\bjtitle{J. Geom. Phys.}
\bvolume{9},
\bfpage{303}--\blpage{368}
(\byear{1992})
{\href{https://arxiv.org/abs/hep-th/9204083}{{arXiv:hep-th/9204083}}}.
\doiurl{10.1016/0393-0440(92)90034-X}
\end{barticle}
\endbibitem

\bibitem{Srisangyingcharoen:2021ndd}
\begin{barticle}
\bauthor{\bsnm{Srisangyingcharoen}, \binits{P.}},
\bauthor{\bsnm{Mansfield}, \binits{P.}}:
\batitle{Effective lagrangian for non-abelian two-dimensional topological field
  theory}.
\bjtitle{Nuclear Physics B}
\bvolume{980},
\bfpage{115798}
(\byear{2022}).
\doiurl{10.1016/j.nuclphysb.2022.115798}
\end{barticle}
\endbibitem

\bibitem{Corradini:2016czo}
\begin{barticle}
\bauthor{\bsnm{Corradini}, \binits{O.}},
\bauthor{\bsnm{Edwards}, \binits{J.P.}}:
\batitle{{Mixed symmetry tensors in the worldline formalism}}.
\bjtitle{JHEP}
\bvolume{05},
\bfpage{056}
(\byear{2016})
{\href{https://arxiv.org/abs/1603.07929}{{arXiv:1603.07929}}}
{[hep-th]}.
\doiurl{10.1007/JHEP05(2016)056}
\end{barticle}
\endbibitem

\bibitem{Edwards:2016acz}
\begin{barticle}
\bauthor{\bsnm{Edwards}, \binits{J.P.}},
\bauthor{\bsnm{Corradini}, \binits{O.}}:
\batitle{{Mixed symmetry Wilson-loop interactions in the worldline formalism}}.
\bjtitle{JHEP}
\bvolume{09},
\bfpage{081}
(\byear{2016})
{\href{https://arxiv.org/abs/1607.04230}{{arXiv:1607.04230}}}
{[hep-th]}.
\doiurl{10.1007/JHEP09(2016)081}
\end{barticle}
\endbibitem

\bibitem{Bastianelli:2015iba}
\begin{barticle}
\bauthor{\bsnm{Bastianelli}, \binits{F.}},
\bauthor{\bsnm{Bonezzi}, \binits{R.}},
\bauthor{\bsnm{Corradini}, \binits{O.}},
\bauthor{\bsnm{Latini}, \binits{E.}},
\bauthor{\bsnm{Ould-Lahoucine}, \binits{K.H.}}:
\batitle{{A worldline approach to colored particles}}.
\bjtitle{J. Phys. Conf. Ser.}
\bvolume{1208}(\bissue{1}),
\bfpage{012004}
(\byear{2019})
{\href{https://arxiv.org/abs/1504.03617}{{arXiv:1504.03617}}}
{[hep-th]}.
\doiurl{10.1088/1742-6596/1208/1/012004}
\end{barticle}
\endbibitem

\bibitem{Samuel:1978iy}
\begin{barticle}
\bauthor{\bsnm{Samuel}, \binits{S.}}:
\batitle{{COLOR ZITTERBEWEGUNG}}.
\bjtitle{Nucl. Phys. B}
\bvolume{149},
\bfpage{517}--\blpage{524}
(\byear{1979}).
\doiurl{10.1016/0550-3213(79)90005-1}
\end{barticle}
\endbibitem

\bibitem{Broda:1995wv}
\begin{botherref}
\oauthor{\bsnm{Broda}, \binits{B.}}:
{NonAbelian Stokes theorem},
496--505
(1995)
{\href{https://arxiv.org/abs/hep-th/9511150}{{arXiv:hep-th/9511150}}}
\end{botherref}
\endbibitem

\bibitem{10.21468/SciPostPhysLectNotes.21}
\begin{botherref}
\oauthor{\bsnm{Haber}, \binits{H.E.}}:
{Useful relations among the generators in the defining and adjoint
  representations of SU(N)}.
SciPost Phys. Lect. Notes,
21
(2021).
\doiurl{10.21468/SciPostPhysLectNotes.21}
\end{botherref}
\endbibitem

\bibitem{PM}
\begin{botherref}
\oauthor{\bsnm{Curry}, \binits{C.}},
\oauthor{\bsnm{Mansfield}, \binits{P.}}:
The Wilson loop for non-Abelian gauge theory as a tensionless string with
  contact interaction.
In preparation
\end{botherref}
\endbibitem

\bibitem{DEAZCARRAGA1998657}
\begin{barticle}
\bauthor{\bsnm{{de Azcárraga}}, \binits{J.A.}},
\bauthor{\bsnm{Macfarlane}, \binits{A.J.}},
\bauthor{\bsnm{Mountain}, \binits{A.J.}},
\bauthor{\bsnm{{Pérez Bueno}}, \binits{J.C.}}:
\batitle{Invariant tensors for simple groups}.
\bjtitle{Nuclear Physics B}
\bvolume{510}(\bissue{3}),
\bfpage{657}--\blpage{687}
(\byear{1998}).
\doiurl{10.1016/S0550-3213(97)00609-3}
\end{barticle}
\endbibitem

\end{thebibliography}


\end{document}